\documentclass[prd,12pt,showpacs,showkeys,aps,nofootinbib]{revtex4-1}
\usepackage[utf8x]{inputenc}
\usepackage[dvips]{graphicx}
\usepackage{amssymb}
\usepackage{amsmath,amssymb,graphicx}
\usepackage{graphicx}
\usepackage{dcolumn}
\usepackage{bm}
\usepackage{fancyhdr}
\usepackage{latexsym}
\usepackage{amsfonts}
\usepackage{amssymb}
\usepackage{amsthm}
\usepackage{amsmath}
\usepackage{color}
\usepackage{colordvi}
\usepackage{array}
\usepackage{hyperref}
\usepackage{slashed}
\usepackage[utf8x]{inputenc}
\usepackage{subfigure}

\def\RCT{R\chi T}

\begin{document}
\title{The $VV'P$ form factors in resonance chiral theory and the $\pi-\eta-\eta'$ light-by-light contribution to the muon $g-2$}
\vskip 6ex
\author{P. Roig}
\affiliation{Instituto de F\'{\i}sica, Universidad Nacional Aut\'onoma de M\'exico,
Apartado Postal 20-364, 01000 M\'exico D.F., M\'exico}
\author{A. Guevara}
\author{G. L\'opez Castro}
\affiliation{Departamento de F\'isica, Centro de Investigaci\'on y de Estudios Avanzados, Apartado Postal 14-740, 07000 M\'exico D.F., M\'exico}

\

\bigskip

\begin{abstract}


The description of the $VV'P$ form factors ($V,V'$ stands for vector particles and $P$ for a pseudoscalar meson) for different particles virtualities remains a challenge for 
the theory of strong interactions. While their chiral limit is well understood, recent measurements of the $\gamma^*\omega\pi^0$ and $\gamma^*\gamma \pi^0$ form factors 
at high photon virtualities seem to depart from the simplest scaling behavior suggested by QCD. Here we attempt to describe them in their whole measured energy regimes 
within the Resonance Chiral Theory, a framework which naturally incorporates the chiral limit constraints and extends to higher energies by including the resonances as 
active fields.

Specifically, we obtained an accurate description of the data up to $2$ GeV on the former form factor by including three multiplets of vector resonances. Good agreement with measurements 
of the latter was possible even in the single resonance approximation, although we propose to measure the $e^+e^-\to\mu^+\mu^-\pi^0$ cross-section and di-muon invariant mass 
distribution to better characterize this form factor. We have then evaluated the pion exchange contribution to the muon $g-2$ obtaining $\left(6.66\pm0.21\right)\cdot 10^{-10}$ 
with an accurate determination of the errors. We have also recalled that approximating the whole pion exchange by the pion pole contribution underestimates the corresponding 
result for the anomaly (by $(15,20)\%$). Based on these results, we have predicted the $\eta^{(\prime)}$ transition form factors obtaining good agreement with data and 
obtained their respective contributions to the muon anomaly. In this way, the contribution of the three lightest pseudoscalars to it yields $\left(10.47\pm0.54\right)\cdot 10^{-10}$, 
in agreement with previous evaluations but with smaller error.


\end{abstract}

\bigskip



\maketitle
\bigskip

\section{Introduction}\label{Intro}

The form factor describing the $\gamma^{(*)}\gamma^{(*)}\pi$ vertex (also called the pion transition form factor, $\pi$TFF) has played an important role in establishing QCD as the 
dynamical theory of strong interactions \cite{BrodskyLepage, Early_Works} and the role of the anomaly for the gauge theory \cite{Chiral_Anomaly}. In the chiral limit, the 
prediction for this vertex \cite{Chiral_pigg} has been beautifully confirmed by the measured rate of $\pi^0 \to \gamma\gamma$ decays \cite{Beringer:1900zz}. The isospin 
related weak vertex $\gamma W^{-*}\pi^+$ has also been proven to obey the chiral limit prediction \cite{radpidec} from measurements of the vector form factor in radiative 
weak decays of pions \cite{Beringer:1900zz}. On the other hand, the QCD predictions for very large photon virtualities \cite{BrodskyLepage} seem to be at odds with recent 
measurements at $B$ factories experiments \cite{Aubert:2009mc, Shen:2013okm}. These predictions for the $\pi$TFF in the infrared and ultraviolet limits have traditionally 
provided a guide to built the vertex in the intermediate energy region, where the effects of hadronic degrees of freedom play a prominent role. This transition energy 
region is particularly relevant for testing the Standard Theory of elementary particles. As a significant example, the evaluation of the hadronic light-by-light (HLbL) 
scattering contribution to the anomalous magnetic moment of the muon ($a_\mu$) is dominated by the pion exchange diagrams which require the $\gamma\gamma\pi$ vertex with all 
the particles off their mass-shells (see \cite{Czyz:2013sga} and references therein~\footnote{The hadronic (vacuum polarization and LbL) contributions to $a_\mu$ are introduced in more 
detail in section \ref{LbL}.}). It is worth to mention that the hadronic contributions to the muon $g-2$ provide the main source of current theoretical uncertainties in the 
Standard Theory prediction for this observable~\footnote{A recent account, with an updated list of references, can be found in Ref.~\cite{Czyz:2013zga}.}, which exhibits a pertinacious 
discrepancy at the three sigma level between the Standard Theory prediction and the BNL measurement \cite{BNL}. This disagreement attracts more attention given the lack of 
new physics signals at the LHC and, together with the forthcoming experiments at Fermilab and J-Parc \cite{Talks} aiming to improve the current uncertainty of $a_\mu$ by a factor four, 
pushes the theoretical community to try to reduce the corresponding theoretical error of $a_\mu$ (which matches the present experimental accuracy). Along these lines, 
decreasing the error of the hadronic vacuum polarization and HLbL scattering contributions to $a_\mu$ turns out to be the main target.

In this paper we use the present experimental information on the related $\gamma^*\omega\pi$ form factor at large photon virtualities to constrain the behavior of the 
transition form factor in the resonance region. The $\gamma^*\omega\pi^0$ interaction has been probed in $\omega\pi^0$ production in electron-positron collisions at energies 
ranging from threshold up to the $\Upsilon(5S)$ center of mass energies \cite{Shen:2013okm, Achasov:2000wy, Achasov:2013btb, Adam:2004pr, Adams:2005ks}. The isospin related 
vertex, $W^{*}\omega\pi^-$, has been measured from threshold up to the tau lepton mass in the $\tau^- \to \omega\pi^-\nu_{\tau}$ decays \cite{Edwards:1999fj}. Previous attempts to 
constraint the short-distance behavior of this form factor \cite{Gerard:1995bv, Gerard:1997ym} relied on theoretical constraints based on the asymptotic behavior predicted by 
QCD on the basis of the BJL theorem \cite{bjl}. Here, we use recent experimental data on the $\gamma^*\omega\pi^0$ form factor in the asymptotic regime as a more realistic 
high-energy constraint to complement the behavior at lower scales. The intermediate energy region of the transition form factor is described in the framework of the Resonance 
Chiral Theory \cite{RChT, Kampf:2011ty} ($R\chi T$) which already incorporates the chiral constraints \cite{ChPT}. The remaining free parameters involved in the form factor 
are fixed from a fit to experimental data for photon virtualities up the $2$ GeV region.

After getting rid of the form factor in the full energy regime covered by current experimental data, we propose to use it to predict the $e^+e^- \to \pi^0\mu^+\mu^-$ 
process~\footnote{Analogous processes involving the $\eta$ and $\eta'$ mesons can be considered as well.} which is driven by the TFF with virtual timelike photons (to the best 
of our knowledge, this process has not been measured yet nor has been studied before)\footnote{The most important piece of information on the the $\pi$TFF is obtained from $\sigma(e^+e^-\to e^+e^-\pi^0)$, measured in a kinematical configuration that singles out the 
$t$-channel contribution which, in turn, can be readily related to the $\pi\gamma^*\gamma$ vertex with good accuracy. We will confront the Resonance Chiral Lagrangian 
prediction to available data on this observable to fix as much as possible the $VV'\pi$ form factor.
}. Conversely, its measurement, which seems to be at the reach of present and forthcoming 
$e^+e^-$ colliders, would provide a valuable information on the $\pi$TFF and an unambiguous test of the $R\chi T$ prediction. As a natural use of our $\gamma^*\gamma^*\pi^0$ 
form factor, we evaluate its contribution to the HLbL piece of the muon $g-2$ paying special attention to a careful evaluation of the associated errors. Using our $\pi$TFF it 
is possible to predict the $\eta$ and $\eta'$ TFF. Their comparison with the data validates our approach and we are thus able to evaluate the corresponding contributions to the 
HLbL scattering muon anomaly. In this way, we obtain the contribution of the three lightest pseudoscalars to the HLbL scattering $g-2$.

The paper is organized as follows: in section \ref{Theory} we explain our theoretical setting, namely $R\chi T$, and introduce the relevant pieces of 
the Lagrangian that will be employed throughout. Next, in section \ref{omegapigammaFF} we present the $R\chi T$ results for the $\gamma^\star \omega \pi$ form factor and discuss 
the QCD short-distance constraints that apply to the involved couplings in section \ref{SD RCT}. In section \ref{Data_omega} we confront this form factor to the 
available data below $2$ GeV obtained from $e^+e^-$ collisions and $\tau$ decays, (section \ref{Data_omega_lowE}) and find that three multiplets of resonances are required to 
obtain good accuracy fits to data, in agreement with Ref.~\cite{Achasov:2013btb}. Our best fit results yield small violations of the high-energy constraints. The observation 
that our best fit form factor does not show a sizable disagreement with data in the charmonium region motivates us to study the possible extension of our description to 
higher energies in section \ref{Data_omega_highE}. We find that, although this is possible up to the $\Psi(2S)$ region, the $R\chi T$ description cannot be extended up to 
larger energies even including an infinite tower of resonances per quantum number, as predicted by large-$N_C$ arguments \cite{LargeNc}. We discuss the possible 
interpretation of this result and extend our $R\chi T$ form factor to higher energies by matching it to a simple ansatz in section \ref{Matching}, in such a way that data in 
the bottomonium region can be accommodated. Unfortunately, we have not been able to find an explanation for these data in the $10$ GeV region~\footnote{Data on the $\pi$TFF only 
extend up to $\sim 6.3$ GeV, so they cannot help settle this issue. Fortunately, the impact of this energy region on $a_\mu^{\pi^0,\,HLbL}$ is completely negligible.}. At this 
point we turn to the $\pi$TFF, whose derivation within $R\chi T$ is recalled in section \ref{piggFF}, where formulae are given both for the virtual and real pion cases. This 
is done by considering, in addition to the pseudo-Goldstone bosons, the lightest multiplet of pseudoscalar and vector resonances, a choice which is motivated by the study of 
consistent short-distance constraints in the odd-intrinsic parity Resonance Chiral Lagrangian \cite{JJYo}. Noteworthy, all involved couplings are predicted in the case with a 
real pion, while only one of them is not in the virtual pion case and needs to be fixed phenomenologically. Data on the $\pi$TFF is analyzed in section \ref{Analysis pig*g}; 
a good agreement with data is found with tiny violations of the asymptotic QCD constraints. In section \ref{probe of pig*g*} we propose the study of a new observable involving 
the $\pi$TFF, namely the $e^+e^-\to \mu^+\mu^-\pi^0$ cross-section and di-muon invariant mass distribution, and discuss the experimental signatures and the feasibility of 
these measurements at present and near future facilities. This reaction can provide complementary information for the $\pi$TFF of timelike photons from its threshold up to 
bottomonium energy scales. As an application of these analyses we evaluate in section \ref{LbL} the dominant pion exchange contribution to the HLbL piece of $a_\mu$ and 
discuss our result confronting it to other predictions in the literature. We also comment on the assumption of considering the pion pole contribution instead of the whole 
pion exchange contribution, which underestimates the result way beyond the quoted errors. The $\eta$ and $\eta^\prime$ exchange contributions to $a_\mu^{HLbL}$ can also be 
computed using our results for the $\pi$TFF, which is done in section \ref{etaetap}. In this way we come up with an evaluation of the leading pseudoscalar exchange 
contributions to $a_\mu^{HLbL}$. Finally, we summarize our findings and present and outlook on $a_\mu^{HLbL}$ in section \ref{Conclusions}. An appendix giving relevant 
formulae for the evaluation of $a_\mu^{P,\,HLbL}$ in sections \ref{LbL} and \ref{etaetap} completes our present work.

\section{Theoretical setting}\label{Theory}

The $\gamma^*\omega\pi^0$ form factor and the $\pi$TFF cannot be derived analytically from first principles since they span diverse energy regions, some of them belonging to 
the non-perturbative regime of QCD. This quantum field theory of the strong interactions predicts the behaviours of these form factors at the two extremes of the energy region. 
On the one hand, the approximate chiral symmetry of light-flavoured QCD yields definite corroborated predictions on the very low energy end. On the other side, perturbative 
QCD allows to derive the asymptotic behaviour of the form factors under study. The resonance chiral Lagrangians are thought as a useful tool to interpolate between these two 
known behaviours.

The effective field theory dual to QCD at low energies is Chiral Perturbation Theory \cite{ChPT} ($\chi PT$), which is based on an expansion in powers of the momenta and/or 
masses of the lightest pseudoscalar mesons (which have status of pseudo-Goldstone bosons of the spontaneous chiral symmetry breakdown) over the chiral symmetry breaking scale, 
of order $1$ GeV. Around this typical hadronic scale, the chiral expansion will no longer be convergent. In fact, $\chi PT$ ceases to be applicable much earlier, at 
$E\lesssim M_\rho$ [with $M_\rho$ the mass of the $\rho(770)$ meson], where new degrees of freedom corresponding to the lightest light-flavoured resonances become active.  

When these resonances are introduced as dynamical fields in the action of the theory the inverse of the number of colours of the $QCD$ gauge group, $1/N_C$ \cite{LargeNc}, 
becomes a useful expansion parameter \cite{Pich:2002xy} for a perturbative approach to the meson resonance dynamics. At leading order (LO) in this expansion the spectrum of the 
theory includes infinite (excited) copies of every meson with definite quantum numbers and these states are free and stable. Next-to-leading order (NLO) corrections explain 
the (rather wide) widths of (many) mesons and their decays by tree-level contact interactions described by an effective Lagrangian.

A realization of these ideas is provided by the Resonance Chiral Theory \cite{RChT, Kampf:2011ty}, $R\chi T$, which is built requiring chiral symmetry for the pseudo-Goldstone 
bosons (spontaneously broken by the quark condensate and explicitly by the small light quark masses), unitary symmetry for the resonance multiplets and the discrete symmetries 
of the QCD Lagrangian, without any dynamical assumption on the role of any type of resonances in the theory. In this respect, the well-known notion of vector meson dominance 
\cite{Sakurai} emerges as a dynamical result \cite{RChT} and not as a priori assumption. The coefficients of the resonance chiral Lagrangians are not restricted by this 
procedure and all of them are free parameters until compatibility with QCD short-distance information is required.

The matching of the $R\chi T$ Green functions and associated form factors to the $QCD$ expressions for these quantities yields restrictions among the resonance couplings that 
ensure a right asymptotic behaviour of the $R\chi T$ expressions and increase the predictability of the theory. Within the antisymmetric tensor formalism, it has been shown 
that a consistent set of short-distance $QCD$ constraints on the $R\chi T$ even- and odd-intrinsic parity couplings can be found including only the lightest multiplet of 
resonances with given quantum numbers \cite{RChT, JJYo}~\footnote{Analogous studies within the vector field formalism were pioneered by Moussallam and Knecht and 
Nyffeler for the $VVP$ Green function \cite{Chiral_pigg}.}. In this way, the minimal hadronic ansatz \cite{MHA} --corresponding to including as many resonance multiplets as 
needed to achieve consistent high-energy constraints on the resonance couplings-- reduces to the single resonance approximation. The discussion of the asymptotic QCD constraints on the $R\chi T$ couplings relevant 
for this work can be found in section \ref{SD RCT}.

It should be pointed out, however, that there is no limitation in $R\chi T$ with respect to the number of meson multiplets to be included in the theory. As a guiding principle, 
the fact that the low-energy dynamics is mostly determined by the lightest states suggests that it is a sound approximation to include only those degrees of freedom that can 
be excited in the considered process, which is the basis of the effective field theory approach. The addition of more resonance multiplets will increase the number of participating 
couplings, reduce the predictability of the theory and modify the short-distance constraints obtained in the single resonance approximation. However, when data are precise enough 
to probe the physics of the excited resonances, these should be added as active fields to the action, as suggested by the $N_C\to\infty$ limit.

The $R\chi T$ Lagrangian relevant for this article is (only the lightest multiplet of pseudoscalar and vector resonances is included, see the discussion below eq.~(\ref{OVVP}) 
for the introduction of a second meson multiplet)
\begin{equation}\label{L_RChT}
 \mathcal{L}_{R\chi T}\,=\,\mathcal{L}_{\chi PT}^{\mathcal{O}(p^2)}\,+\,\mathcal{L}_{WZW}^{\mathcal{O}(p^4)}\,+\,{\cal L}^{{ \rm kin},\,R}_{R \chi T}\,+\,\mathcal{L}_{R\chi T}^{V}\,+
\,\mathcal{L}_{R\chi T}^{P}\,+\,\mathcal{L}_{R\chi T}^{VJP}\,+\,\mathcal{L}_{R\chi T}^{VVP}\,+\,\mathcal{L}_{R\chi T}^{P,\,\mathrm{rest}}\,,
\end{equation}
where
\begin{equation}\label{L_ChPT_p2}
 \mathcal{L}_{\chi PT}^{\mathcal{O}(p^2)}\,=\,\frac{F^2}{4}\left\langle u_\mu u^\mu+\chi_+\right\rangle
\end{equation}
is the lowest order $\chi PT$ Lagrangian, with
\begin{eqnarray}
u_{\mu} & = & i [ u^{\dagger}(\partial_{\mu}-i r_{\mu})u-
u(\partial_{\mu}-i \ell_{\mu})u^{\dagger} ] \ , \nonumber \\
\chi_{\pm} & = & u^{\dagger}\chi u^{\dagger}\pm u\chi^{\dagger} u\ \
\ \ , \ \ \ \
\chi=2B_0(s+ip) \; \; ,
\end{eqnarray}
and $\langle \ldots \rangle$ is short for a trace in the flavour space. The pseudo-Goldstone nonet of pseudoscalar fields is realized non-linearly into the unitary 
matrix $u$ (which includes the familiar exponential of the matrix with the $\pi$, $K$ and $\eta^{(\prime)}$ meson fields) in the flavour space and the external hermitian 
fields $s$, $p$, $\ell^\mu$ and $r^\mu$ promote the global chiral symmetry to a local one, enabling the introduction of the electroweak interactions (through $\ell^\mu$ and 
$r^\mu$) and the explicit chiral symmetry breaking (via $s$) by means of these auxiliary fields (see Refs.~\cite{RChT} for details). $F$ (the pion decay constant) and $B_0$ 
(related to the quark condensate) are the two lowest order $\chi PT$ low-energy constants in the chiral limit.

The leading action in the odd-intrinsic parity Lagrangian is given by the chiral anomaly of QCD, which is explicitly fulfilled by the Wess-Zumino-Witten \cite{WZW} functional that can 
be read in Ref.~\cite{Ecker:1994gg}. $\mathcal{L}_{WZW}^{\mathcal{O}(p^4)}$ contains all anomalous contributions to electromagnetic and semileptonic meson decays at order 
$\mathcal{O}(p^4)$ in the chiral expansion.

The terms $\mathcal{L}_{\chi PT}^{\mathcal{O}(p^2)}\,+\,\mathcal{L}_{WZW}^{\mathcal{O}(p^4)}$ make evident that the $R\chi T$ Lagrangian, eq.~(\ref{L_RChT}), reproduces by 
construction the LO $\chi PT$ Lagrangian both in the odd- ($\mathcal{O}(p^4)$) and even-intrinsic ($\mathcal{O}(p^2)$) parity sectors. The use of the antisymmetric tensor 
representation for the spin-one fields is convenient because the NLO chiral LECs (in both sectors) are saturated upon integration of the resonance fields. Therefore, the 
$\chi PT$ Lagrangian at NLO in both parity sectors does not have to be included in eq.~(\ref{L_RChT}) to avoid double counting. Contributions generated by loops including 
$\mathcal{L}_{\chi PT}^{\mathcal{O}(p^2)}\,+\,\mathcal{L}_{WZW}^{\mathcal{O}(p^4)}$, which are NLO in both sectors, can be included through the off-shell meson widths~\cite{GomezDumm:2000fz}, 
requiring analyticity \cite{Realpartwidths} and by the renormalization procedure in $\chi PT$.

The 'kinetic' terms (which also include interactions bilinear in the resonance fields through the covariant derivative) for the resonances, ${\cal L}^{{ \rm kin},\,R}_{R \chi T}$, 
can be found in Refs.~\cite{RChT}. The resonance chiral Lagrangians with one vector or pseudoscalar resonance nonet and an $\mathcal{O}(p^2)$ chiral tensor are
\begin{equation}
{\cal L}_{R\chi T}^V \, = \,  \frac{F_V}{2\sqrt{2}} \langle V_{\mu\nu} f_+^{\mu\nu}\rangle + i\,\frac{G_V}{\sqrt{2}} \langle V_{\mu\nu} u^\mu u^\nu\rangle\,,\quad
{\cal L}_{R\chi T}^P \, = \, i\,d_m \left\langle P \chi_-\right\rangle\,,
\end{equation}
where
$f_\pm^{\mu\nu}  =  u F_L^{\mu\nu} u^\dagger \pm u^\dagger F_R^{\mu\nu}
u$ and $F_{R,L}^{\mu \nu}$ are the field strength tensors associated
with the external right- and left-handed auxiliary fields. Hereafter, all couplings will be taken as real parameters.

The odd-intrinsic parity resonance Lagrangian with two vector objects and an $\mathcal{O}(p^2)$ chiral tensor is written (here $P$ stands for a pseudoscalar meson, following 
the notation of Ref.~\cite{RuizFemenia:2003hm})
\begin{equation}\label{L RChT odd}
\mathcal{L}_{R\chi T}^{VJP}\,+\,\mathcal{L}_{R\chi T}^{VVP}\,=\,\sum_{i=1}^{7}c_i\,\mathcal{O}_{VJP}^i\,+\,\sum_{j=1}^{4}d_i\,\mathcal{O}_{VVP}^j\,,
\end{equation}
in terms of the following operators~\footnote{The Lagrangian in eq.~(\ref{L RChT odd}) is complete for constructing vertices with only pseudoscalar~\cite{RuizFemenia:2003hm}.}
\begin{eqnarray}
 {\cal O}_{VJP}^1 & = & \varepsilon_{\mu\nu\rho\sigma}\,
\langle \, \{V^{\mu\nu},f_{+}^{\rho\alpha}\} \nabla_{\alpha}u^{\sigma}\,\rangle
\; \; , \nonumber\\[2mm]
{\cal O}_{VJP}^2 & = & \varepsilon_{\mu\nu\rho\sigma}\,
\langle \, \{V^{\mu\alpha},f_{+}^{\rho\sigma}\} \nabla_{\alpha}u^{\nu}\,\rangle
\; \; , \nonumber\\[2mm]
{\cal O}_{VJP}^3 & = & i\,\varepsilon_{\mu\nu\rho\sigma}\,
\langle \, \{V^{\mu\nu},f_{+}^{\rho\sigma}\}\, \chi_{-}\,\rangle
\; \; , \nonumber\\[2mm]
{\cal O}_{VJP}^4 & = & i\,\varepsilon_{\mu\nu\rho\sigma}\,
\langle \, V^{\mu\nu}\,[\,f_{-}^{\rho\sigma}, \chi_{+}]\,\rangle
\; \; , \nonumber\\[2mm]
{\cal O}_{VJP}^5 & = & \varepsilon_{\mu\nu\rho\sigma}\,
\langle \, \{\nabla_{\alpha}V^{\mu\nu},f_{+}^{\rho\alpha}\} u^{\sigma}\,\rangle
\; \; ,\nonumber\\[2mm]
{\cal O}_{VJP}^6 & = & \varepsilon_{\mu\nu\rho\sigma}\,
\langle \, \{\nabla_{\alpha}V^{\mu\alpha},f_{+}^{\rho\sigma}\} u^{\nu}\,\rangle
\; \; , \nonumber\\[2mm]
{\cal O}_{VJP}^7 & = & \varepsilon_{\mu\nu\rho\sigma}\,
\langle \, \{\nabla^{\sigma}V^{\mu\nu},f_{+}^{\rho\alpha}\} u_{\alpha}\,\rangle
\;\; ,
\label{eq:VJP}
\end{eqnarray}
and
\begin{eqnarray}
{\cal O}_{VVP}^1 & = & \varepsilon_{\mu\nu\rho\sigma}\,
\langle \, \{V^{\mu\nu},V^{\rho\alpha}\} \nabla_{\alpha}u^{\sigma}\,\rangle
\; \; , \nonumber\\[2mm]
{\cal O}_{VVP}^2 & = & i\,\varepsilon_{\mu\nu\rho\sigma}\,
\langle \, \{V^{\mu\nu},V^{\rho\sigma}\}\, \chi_{-}\,\rangle
\; \; , \nonumber\\[2mm]
{\cal O}_{VVP}^3 & = & \varepsilon_{\mu\nu\rho\sigma}\,
\langle \, \{\nabla_{\alpha}V^{\mu\nu},V^{\rho\alpha}\} u^{\sigma}\,\rangle
\; \; , \nonumber\\[2mm]
{\cal O}_{VVP}^4 & = & \varepsilon_{\mu\nu\rho\sigma}\,
\langle \, \{\nabla^{\sigma}V^{\mu\nu},V^{\rho\alpha}\} u_{\alpha}\,\rangle
\; \; .
\label{eq:VVP}
\end{eqnarray}
An equivalent basis for the operators in eq.~(\ref{L RChT odd}) was given in Ref.~\cite{Kampf:2011ty}. The relations between both operator basis can be found in Ref.~\cite{JJYo}.

The remaining part of the odd-intrinsic parity Lagrangian involving pseudoscalar and vector resonances and chiral tensors was derived in Ref.~\cite{Kampf:2011ty}.
We rewrite it as
\begin{equation}\label{L RChT P KN}
 \mathcal{L}_{R\chi T}^{P,\,\mathrm{rest}}\,=\,\sum_{i=1}^{5}\kappa_i^P\varepsilon_{\mu\nu\alpha\beta}\mathcal{O}_i^{P\,\mu\nu\alpha\beta}\,+\,\sum_{j=1}^{3}\kappa_j^{PV}
\varepsilon_{\mu\nu\alpha\beta}\mathcal{O}_j^{PV\,\mu\nu\alpha\beta}\,+\,\varepsilon_{\mu\nu\alpha\beta} \mathcal{O}^{VVP\,\mu\nu\alpha\beta}\,,
\end{equation}
where
\begin{eqnarray}
 \mathcal{O}_1^{P\,\mu\nu\alpha\beta} & = &  \left\langle P\left\lbrace f_-^{\mu\nu},f_-^{\alpha\beta}\right\rbrace \right\rangle \; , \nonumber\\[2mm]
 \mathcal{O}_2^{P\,\mu\nu\alpha\beta} & = &  i \left\langle P u^\alpha f_+^{\mu\nu} u^\beta \right\rangle \; , \nonumber\\[2mm]
 \mathcal{O}_3^{P\,\mu\nu\alpha\beta} & = &  i \left\langle P \left\lbrace f_+^{\mu\nu},u^\alpha u^\beta\right\rbrace\right\rangle \; , \nonumber\\[2mm]
 \mathcal{O}_4^{P\,\mu\nu\alpha\beta} & = &  \left\langle P u^\mu u^\nu u^\alpha u^\beta\right\rangle \; , \nonumber\\[2mm]
 \mathcal{O}_5^{P\,\mu\nu\alpha\beta} & = &  \left\langle P\left\lbrace f_+^{\mu\nu},f_+^{\alpha\beta}\right\rbrace\right\rangle \; ; 
\label{O_i^P}
\end{eqnarray}

\begin{eqnarray}
 \mathcal{O}_1^{PV\,\mu\nu\alpha\beta} & = &  i \left\langle \left\lbrace V^{\mu\nu},P\right\rbrace u^\alpha u^\beta\right\rangle \; , \nonumber\\[2mm]
 \mathcal{O}_2^{PV\,\mu\nu\alpha\beta} & = &  i \left\langle V^{\mu\nu} u^\alpha P u^\beta \right\rangle \; , \nonumber\\[2mm]
 \mathcal{O}_3^{PV\,\mu\nu\alpha\beta} & = &   \left\langle \left\lbrace V^{\mu\nu},P\right\rbrace f_+^{\alpha\beta}\right\rangle \; , 
 \label{O_i^PV}
\end{eqnarray}
and
\begin{equation}
\mathcal{O}^{VVP\,\mu\nu\alpha\beta}\,=\,\left\langle V^{\mu\nu} V^{\alpha\beta} P\right\rangle\,.
\label{OVVP}
\end{equation}

If additional heavier meson multiplets are required by the data, in addition to the replication of the previous Lagrangian for the corresponding excited multiplet (see 
Ref.~\cite{Mateu:2007tr}) there will be additional operators too. For those with two vector fields each of them can belong to a different multiplet giving rise to new terms. 
This was worked out for the $VV'P$ terms in Ref.~\cite{Mateu:2007tr} obtaining
\begin{equation}\label{VV'P}
 \mathcal{L}_{R\chi T}^{V_1 V_2}\,=\,\sum_{n=a,...,e}d_n\, \varepsilon_{\mu\nu\alpha\beta} \,\mathcal{O}^{VV'P\, \mu\nu\alpha\beta}_n\,
,
\end{equation}
with
\begin{eqnarray}
 \mathcal{O}^{VV'P\, \mu\nu\alpha\beta}_a & = &\left\langle \left\lbrace V_1^{\mu\nu},V_2^{\alpha\rho}\right\rbrace \nabla_\rho u^\beta\right\rangle\,,\nonumber\\
 \mathcal{O}^{VV'P\, \mu\nu\alpha\beta}_b & = &\left\langle \left\lbrace V_1^{\mu\rho},V_2^{\alpha\beta}\right\rbrace \nabla_\rho u^\nu\right\rangle\,,\nonumber\\
 \mathcal{O}^{VV'P\, \mu\nu\alpha\beta}_c & = &\left\langle \left\lbrace \nabla_\rho V_1^{\mu\nu},V_2^{\alpha\rho}\right\rbrace u^\beta \right\rangle\,,\nonumber\\
 \mathcal{O}^{VV'P\, \mu\nu\alpha\beta}_d & = &\left\langle \left\lbrace \nabla_\rho V_1^{\mu\rho},V_2^{\alpha\beta}\right\rbrace u^\nu\right\rangle\,,\nonumber\\
 \mathcal{O}^{VV'P\, \mu\nu\alpha\beta}_e & = &\left\langle \left\lbrace \nabla^\beta V_1^{\mu\nu},V_2^{\alpha\rho}\right\rbrace u_\rho\right\rangle\,
.
\end{eqnarray}

\section{$\omega\pi\gamma^\star$ form factor}\label{omegapigammaFF}

The cross-section for the $\omega\pi^0$ production in electron-positron collisions can be written as~\cite{Gerard:1997ym} 
\begin{equation}\label{eq:sigma}
 \sigma\left(e^+e^-\to\omega\pi^0\right)\,=\,\frac{\pi\alpha^2}{6s^3}\lambda^{3/2}\left(s,M_\omega^2,m_\pi^2\right)|F_V^{\omega\pi^0}(s)|^2\,,
\end{equation}
with $\lambda(a,b,c)=a^2+b^2+c^2-2ab-2ac-2bc$. The electromagnetic hadronic form factor is defined from 
\begin{equation}\label{definition matrix element}
\langle \omega(q,\epsilon)\pi^0(p) |\sum_qe_q\bar{q}\gamma_{\mu}q|0\rangle =eF_V^{\omega\pi^0}(s)\varepsilon_{\mu\nu\alpha\beta}(p+q)^{\nu}\epsilon^{\alpha}q^{\beta}\,,
\end{equation}
where $s=(p+q)^2$ is the center-of-mass energy squared. An isospin rotation of this isovector hadronic matrix element \footnote{$G$-parity forbids an isoscalar contribution to $e^+e^- \to \omega\pi^0$. Thus, only the isovector $(I=1)$ part of the electromagnetic current contributes.} allows to define the $F_V^{\omega\pi^-}(s)$ form factor 
which drives the $\tau^-\to\omega\pi^-\nu_\tau$ decay. This decay has been studied in Ref.~\cite{Guo:2008sh} (see also Refs.~\cite{Terschluesen:2010ik, Schneider:2012ez} for 
other studies of this form factor focusing at lower energies), where it was concluded that a sensible description of the corresponding data for the vector spectral function  was possible only 
by including two multiplets of vector resonances in the spectrum of the theory \footnote{This and other related analyses have been updated and refined very recently, 
see Ref.~\cite{Chen:2013nna} for details.}. In complete analogy to Ref.~\cite{Guo:2008sh}, the relevant form factor reads~\footnote{In case a third multiplet of resonances is 
required by the data, the term $-\frac{2F_{V_2}}{FM_\omega}\left(\tilde{d}_m m_\pi^2+\tilde{d}_M M_\omega^2+\tilde{d}_s s\right)D_{\rho^{\prime\prime}}(s)$ 
should be added to eq.~(\ref{eq:FF}). The new couplings $F_{V_2}$, $\tilde{d}_{m,M,s}$ are defined in analogy to the respective $\rho'=\rho(1450)$ couplings.}
\begin{eqnarray}\label{eq:FF}
 F_V^{\omega\pi^0}(s)&=&\frac{2\sqrt{2}}{FM_VM_\omega}\left(c_{1235}m_\pi^2-c_{1256}M_\omega^2+c_{125}s\right)-\frac{4F_V}{FM_\omega}\left[d_{123}m_\pi^2+d_3(s+M_\omega^2)\right]D_\rho(s)\nonumber\\
&-&\frac{2F_{V_1}}{FM_\omega}\left(d_m m_\pi^2+d_M M_\omega^2+d_s s\right)D_{\rho^\prime}(s)\,,
\end{eqnarray}
where the following combinations of couplings --in terms of those in eqs.~(\ref{L RChT odd}) and (\ref{VV'P})--  were defined~\cite{Guo:2008sh, RuizFemenia:2003hm, Dumm:2009kj}
\begin{eqnarray}\label{eq:definition_couplings}
c_{1235} & = & c_1+c_2+8c_3-c_5\,,\;\; c_{1256}=c_1-c_2-c_5+2c_6\,,\;\; c_{125}=c_1-c_2+c_5\,,\\
d_{123}  & = & d_1+8d_2-d_3\,,\;\; d_m=d_a+d_b-d_c+8d_f\,,\;\; d_M=d_b-d_a+d_c-2d_d\,,\;\; d_s=d_c+d_a-d_b\,.\nonumber
\end{eqnarray}
Along the present study we will always assume ideal mixing for the $\omega$ and $\phi$ mesons. Departures from this scheme have been studied in Ref.~\cite{Chen:2013nna} for 
$\tau^-\to\omega\pi ⁻\nu_\tau$ and other related decays.

The resonant shape factors in eq.~(\ref{eq:FF}) are
\begin{equation} \label{D_R(x)}
 D_R(x)\,=\,\frac{1}{M_R^2-x-iM_R\Gamma_R(x)}\,.
\end{equation}
Since the $\rho(770)$ and $\rho(1450)$ states are rather wide resonances, the energy dependence of their decay widths becomes relevant. In the case of the $\rho(770)$ meson 
this question has been studied within the theory yielding \cite{GomezDumm:2000fz}
\begin{equation}
 \Gamma_\rho(s)\,=\,\frac{sM_V}{96\pi F^2}\left[\sigma_\pi^3(s)\theta\left(s-4m_\pi^2\right)+\frac{1}{2}\sigma_K^3(s)\theta\left(s-4m_K^2\right)\right]\,,
\end{equation}
where $\sigma_P(s)=\sqrt{1-4m_P^2/s}$. In this way, the on-shell $\rho(770)$ width is fixed in terms of the resonance mass and known couplings. On the contrary, there is no 
guidance from the chiral limit that applies to the $\rho(1450)$ case (and, eventually, to higher excitations). For simplicity, we will assume its off-shell width is given 
by \cite{Dumm:2009va}
\begin{equation}
 \Gamma_{\rho^\prime}(s)\,=\,\Gamma_{\rho^\prime}\frac{s}{M_{\rho^\prime}^2}\frac{\sigma_\pi^3(s)}{\sigma_\pi^3(M_{\rho^\prime}^2)}\theta\left(s-4m_\pi^2\right)\,,
\end{equation}
with the mass(es) and on-shell width(s) of the $\rho$-like resonance(s) as given by the PDG \cite{Beringer:1900zz}. Even though this does not need to be the case, we anticipate 
that the good agreement found with data does not seem to require them to be free parameters. 

\subsection{Short-distance constraints on the $\RCT$ couplings}\label{SD RCT}

Although a Brodsky-Lepage \cite{BrodskyLepage}, $\sim s^{-1}$, asymptotic behaviour is usually demanded to the $\pi$TFF at high energies, this may be argued~\footnote{It 
is known that the imaginary part of the spin-one correlators goes to a constant value at infinite momentum transfer when evaluated at the parton level \cite{Floratos:1978jb}. 
A local duality interpretation usually leads to the assumption that every one of the infinite number of form factors contributing to these spectral functions must vanish at 
infinite momentum transfer.}. The $e^+e^-\to \omega\pi^0$ cross-section data collected by Belle \cite{Shen:2013okm} in the $\Upsilon(4S)-\Upsilon(5S)$ region cast doubts on 
the validity of the Brodsky-Lepage \cite{BrodskyLepage} conditions for the $F^{\omega \pi^0}_V(s)$ form factor (assuming $\sim10.5$ GeV is a high enough asymptotic energy 
scale). In particular, data fall faster than the Brodsky-Lepage prediction, approaching the $s^{-2}$ behaviour in the bottomonium region~\cite{Shen:2013okm}. Since the Brodsky-Lepage conditions do not 
seem to be enough to warrant a proper asymptotic behaviour, the use of accurate high-energy data to determine the remaining free couplings seems the only alternative to solve 
the puzzle. We assume that they are necessary (but not sufficient) conditions to meet the asymptotics hinted by Belle's data \cite{Shen:2013okm}.

Therefore, we shall start demanding the Brodsky-Lepage behaviour to the form factor in eq.~(\ref{eq:FF}). The relations obtained from this condition must be (and are) compatible with those found by  
studying the two- and three-point Green functions and associated form factors within $\RCT$ \cite{RChT, Kampf:2011ty, RuizFemenia:2003hm, Cirigliano:2004ue, Cirigliano:2005xn, 
Mateu:2007tr, Guo:2008sh, Dumm:2009va, Dumm:2009kj}. In agreement with general field theory considerations, a consistent set of high-energy constraints can be found for all 
these processes \cite{RChT, JJYo}. Those that play a role in our study are
\begin{eqnarray}\label{SD}
 F_V\,=\,\sqrt{3}F\,,\quad c_{125}\,=\,0\,,\quad c_{1256}\,=\,-\frac{N_C M_V}{32\sqrt{2}\pi^2F_V}\sim-3.26\cdot10^{-2}\,,\quad c_{1235}\,=\,0\,,\nonumber\\
 d_{123}\,=\,\frac{F^2}{8F_V^2}=\frac{1}{24}\,,\quad d_3\,=\,-\frac{N_C M_V^2}{64\pi^2F_V^2}\sim-0.112\,,\quad d_s\,=\,\frac{\sqrt{2}M_Vc_{1256}-2d_3F_V}{F_{V_1}}=0\,,\quad
\end{eqnarray}
where the last equation was obtained demanding a Brodsky-Lepage behaviour to $F_V^{\omega\pi^0}(s)$. We point out that the short-distance large-$N_C$ predictions for $d_3$ 
and $d_s$ were not realized in Refs.~\cite{Guo:2008sh, Chen:2013nna}, where they were assumed to be free parameters. Our constraint is the first one showing
that $d_s$ vanishes in the $N_C\to\infty$ limit \footnote{This result has been obtained neglecting the effect of the third and higher vector multiplets, 
otherwise the condition reads $d_s F_{V_1}+\tilde{d_s} F_{V_2}+...=0$.}. The results obtained for $d_3$ in Refs.~\cite{Guo:2008sh, Chen:2013nna} are roughly twice as large as 
our prediction, while their values for $d_s$, $d_s\in\left[-0.32,-0.08\right]$, are non-vanishing.

For later use, we point that the above condition for $d_3$ is equivalent to~\cite{JJYo}
\begin{equation}\label{equivalent_SD_d3}
d_3\,=\,-\frac{N_C}{64\pi^2}\frac{M_V^2}{F_V^2}\,+\,\frac{F^2}{8F_V^2}\,+\,\frac{4\sqrt{2}P_2}{F_V}\,,
\end{equation}
provided the pseudoscalar resonance coupling $P_2\equiv d_m\kappa_3^{PV}$ fulfills~\cite{Kampf:2011ty}
\begin{equation}\label{consistency condition for P_2}
P_2=-\frac{F^2}{32\sqrt{2}F_V}=-\frac{F}{32\sqrt{6}}\,,
\end{equation}
which belongs to the consistent set of short-distance constraints on the odd-intrinsic parity $R\chi T$ couplings\cite{JJYo}. The vanishing of $P_1\equiv d_m\kappa_5^{P}$ is also derived in Ref~\cite{Kampf:2011ty} by requiring the matching of the $\left\langle VVP\right\rangle$ $R\chi T$ Green function 
to the corresponding OPE result.

The parameters in eq. (\ref{eq:FF}) that remain free after applying the short-distance constraints, eqs.~(\ref{SD}), will be fixed from the fit to available data on 
$F_V^{\omega\pi^0}(s)$ in the resonance region. This is our departing strategy which will be slightly modified from the goodness of the fit requirement.

\section{Comparison to experimental data on the $\omega\pi$ form factor}\label{Data_omega}

\subsection{Data below 2 GeV}\label{Data_omega_lowE}

In the energy region below the $\tau$ lepton mass, the $\gamma^*\omega\pi$ form factor can be obtained either from $\tau$ decays or $e^+e^-$ annihilations. The vector spectral function can be 
extracted from the $\tau^-\to \pi^-\pi^-\pi^+\pi^0\nu_\tau$ decays measured by CLEO \cite{Edwards:1999fj}, by isolating the  $\omega(782)$ meson contribution from its 
three-pion decay channel. This was done by rescaling the $3\pi$ invariant mass distribution with the $BR(\omega\to\pi^+\pi^-\pi^0)=(89.2\pm0.7)\%$ fraction \cite{Beringer:1900zz}.
A complementary piece of information in this energy region comes from the $e^+e^-\to \pi^0 \pi^0\gamma$ cross-section, which has been measured with good precision by 
SND \cite{Achasov:2000wy, Achasov:2013btb} from threshold up to $2$ GeV \footnote{The KLOE Collaboration \cite{Ambrosino:2008gb} measured this observable in a window of $30$ 
MeV around the $\phi(1020)$ meson peak, where the effects of this resonance pop up through the interference with the dominant $\omega(782)$ meson contribution. Since the 
study of this interference is not among our purposes, we will not consider these data in our analysis.}. Since these final states are dominated by the intermediate 
$\omega(782)$ resonance, one needs to rescale the data by the $BR(\omega\to\pi^0\gamma)=(8.28\pm0.28)\%$ in order to obtain the $\omega \pi^0$ cross-section~\footnote{The 
energy-dependence of the $\omega(782)$ decay can be neglected. This assumption is supported by the CVC analysis of Ref.~\cite{Achasov:2013btb}, which shows a nice agreement 
of the $\omega\pi$ data produced in $\tau$ and $e^+e^-$. We thank Leonid Kardapoltsev for conversations on this topic.}.

In order to fit the SND and CLEO data using the form factor in eq.~(\ref{eq:FF}), tau decay data needs to be isospin-rotated to obtain the $e^+e^-$ 
cross-section. In terms of the vector spectral function, $V_{\omega\pi^-}(s)$ measured in tau decays, and rescaling it by the three-pion branching fraction of the $\omega(782)$ 
meson decay, one gets:
\begin{equation}
 \sigma_{\omega\pi^0}(s)\,=\,\frac{4\pi^2\alpha^2}{s}V_{\omega\pi^-}(s)\frac{100}{BR\left(\omega\to\pi^+\pi^-\pi^0\right)}\,.
\end{equation}

After the consistent set of short-distance constraints (\ref{SD}) has been imposed to eq. (\ref{eq:FF}), this only depends on three unknown couplings: $F_{V_1}$, $d_m$ and $d_M$. 
Floating these couplings  is not sufficient to obtain a good agreement with data and the introduction of the third multiplet of resonances does not have enough influence to 
change this result. We understand this because the relations (\ref{SD}) have an associated error of order one third. Therefore, we shall allow to vary the couplings involved 
in the last of these relations \footnote{Since many phenomenological studies have described data accurately using the remaining relations \cite{RuizFemenia:2003hm, Mateu:2007tr, 
Guo:2008sh, Dumm:2009kj, Guo:2010dv, Kampf:2011ty, Dumm:2012vb, Chen:2012vw, Dai:2013joa} we will stick to the values in eqs.~(\ref{SD}) for them.} in an according range.

This increase in the number of fitted couplings makes the fits unstable. From our previous study we have noticed that $d_m$ of order unity influences $\sigma(e^+e^-\to\omega\pi^0)$ 
very slightly. We will fix it to $-1$ for definiteness, as done in Ref.~\cite{Guo:2008sh}. Also, since the coupling $F_{V_1}$ can be determined rather accurately \cite{Guo:2008sh} 
we will follow this evaluation and set it to $(-0.10\pm0.01)$ GeV. Variations within the error do not affect substantially the results. Within this setting, we will first 
consider only the contributions from the first and second resonance multiplets only and then will treat the addition of the heavier excitations.

\begin{figure}[h!]
\begin{center}
\vspace*{1.25cm}
\includegraphics[scale=0.5,angle=-90]{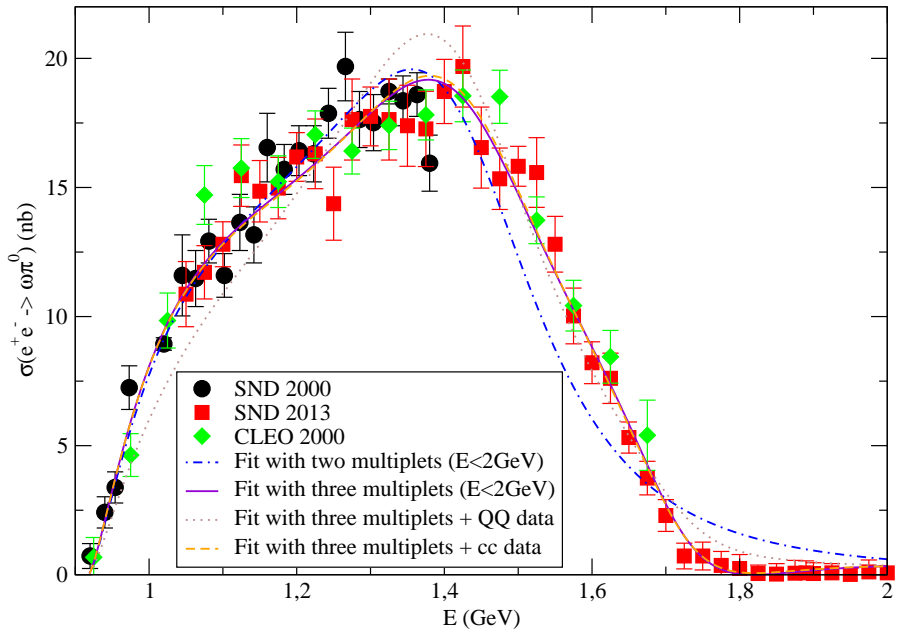}
\caption[]{\label{fig:Fit_low} \small{SND \cite{Achasov:2000wy, Achasov:2013btb} and isospin-rotated CLEO data \cite{Edwards:1999fj} for $\sigma(e^+e^-\to \omega \pi^0)$ are 
confronted to the best fit results including two (blue dashed-dotted line) or three resonance multiplets (solid purple line). Fits obtained including data in the charmonium 
region (orange dashed line) as well as in the bottomonium region (grey dotted line) are also displayed. Higher-energy data can be seen in Fig.~\ref{fig:Fit_high}.}}
\end{center}
\end{figure}

Under the above approximations, the $e^+e^-\to\omega\pi^0$ cross section depends on four unknown coupling constants: $c_{1256}$, $d_3$, $d_M$ and $d_s$. We have fitted them to SND and CLEO data obtaining
\begin{equation}\label{eq:Fit1}
 c_{1256}=-0.037\pm0.002\,,\quad d_3=-0.174\pm0.004\,,\quad d_M=0.41\pm0.09\,,\quad d_s=-0.27\pm0.02\,,
\end{equation}
with $\chi^2/ndf=3.9$. This fit is represented by the blue dashed-dotted line in Fig.~\ref{fig:Fit_low}. While the description of the data is quite good in the lower half of 
the spectrum, disagreement is seen for $\sqrt{s}>1.5$ GeV, which seems to require the contribution of a third multiplet of vector resonances to achieve a better description of the data. This 
feature is not visible in tau data according to the analysis of Refs.~\cite{Guo:2008sh, Chen:2013nna} because using only tau data there are scarcely three points in this region 
with clearly larger errors than in the electron-positron cross-section. We note that these numerical values violate the relations for $c_{1256}$ and $d_3$ in eqs.~(\ref{SD}) 
very little, although the violation for the last relation in eq. (\ref{SD}) is around $50\%$.

Next we have included a third multiplet of resonances, as indicated in footnote $\left[1\right]$. As in the previous case, we will assume $\tilde{d}_m=-1$ so that we are 
introducing three new free parameters: $\tilde{F}_{V_1}$, $\tilde{d}_M$ and $\tilde{d}_s$. The best fit result to SND and CLEO data yields 
\begin{eqnarray}\label{eq:Fit2}
 c_{1256} & = & -0.055\pm0.004\,,\quad d_3=-0.180\pm0.005\,,\quad d_M=0.86\pm0.12\,,\quad d_s=-0.33\pm0.04\,,\nonumber\\
 \tilde{F}_{V_1} & = & 0.079\pm0.004\,,\quad \tilde{d}_M=\,2.05\pm0.17,\quad \tilde{d}_s=-0.42\pm0.04\,,
 \end{eqnarray}
with $\chi^2/ndf=1.2$. The corresponding curve is shown as a solid purple line in Fig.~\ref{fig:Fit_low}, where good agreement with measurements can be appreciated in the 
whole data range. The violations of the short distance constraints for $c_{1256}$ and $d_3$ continue to be reasonable and the two terms 
$F_{V_1}d_s+\tilde{F}_{V_1}\tilde{d}_s$ compensate enough each other to yield a small violation of the last short-distance relation which is within expectations.

Small variations in the values  of the couplings not fitted in eqs.~(\ref{eq:Fit2}) should yield a reasonable estimate of our systematic error. We have found that this uncertainty estimate is 
completely dominated by the value of $F_V$. $F_V=\sqrt{3}F$ has been obtained in a variety of analyses~\cite{RChT, Kampf:2011ty, Dumm:2009kj, Pich:2010sm, Nieves:2011gb, 
Guo:2012ym, Guo:2012yt}. In the context of tau decays, the TAUOLA determination of this parameter, fitting three- and two-pion invariant mass distributions in 
$\tau^-\to\pi^+\pi^-\pi^-\nu_\tau$ decays \cite{Shekhovtsova:2012ra, Nugent:2013hxa}, is compatible within the $5\%$ quoted error with this result and does not indicate a 
sizable deviation due to the presence of excited resonance multiplets. Since its value will be the dominant source of systematic error, we will consider 
a $10\%$ variation around this prediction to estimate conservatively our errors. 

\subsection{Data above 2 GeV: an approach with heavier resonances }\label{Data_omega_highE}

\begin{figure}[h!]
\begin{center}
\vspace*{1.25cm}
\includegraphics[scale=0.5,angle=-90]{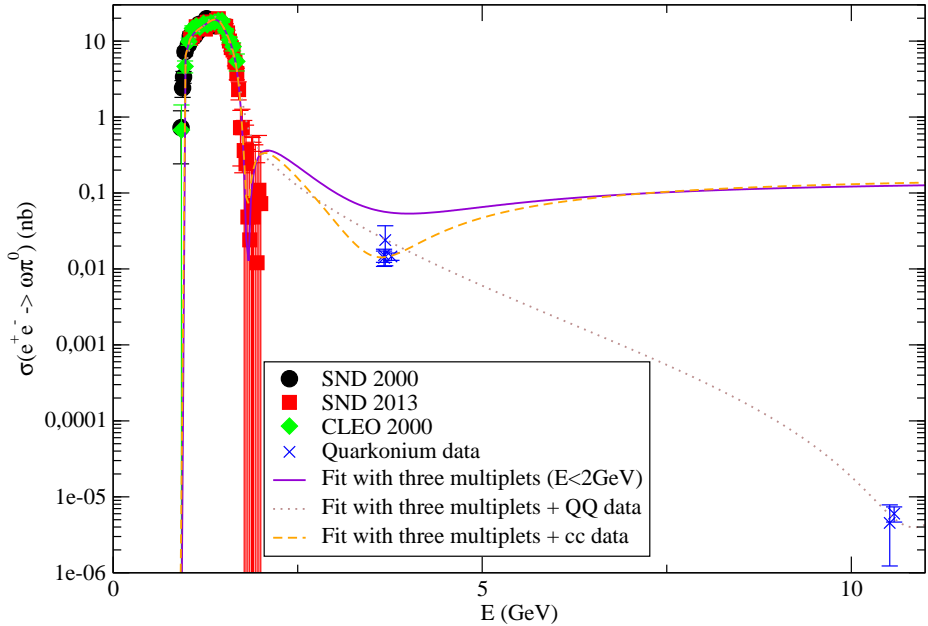}
\caption[]{\label{fig:Fit_high} \small{SND \cite{Achasov:2000wy, Achasov:2013btb}, isospin-rotated CLEO data \cite{Edwards:1999fj}, CLEO data in the charmonium region 
\cite{Adam:2004pr, Adams:2005ks} and Belle \cite{Shen:2013okm} data in the bottomonium region for $\sigma(e^+e^-\to \omega \pi^0)$ are confronted to the best fit results including three resonance multiplets (solid purple 
line) as well as data in the charmonium region (orange dashed line) and also in the bottomonium region (grey dotted line).}}
\end{center}
\end{figure}

A few experimental data points for the $\omega\pi^0$ form factor above 2 GeV have been obtained from electron-positron annihilation experiments. In Fig.~\ref{fig:Fit_high} we 
have included these data points: three of them have been obtained using CLEO Collaboration data in the $\psi$ region \cite{Adam:2004pr, Adams:2005ks} and three more of them 
were very recently reported by Belle Collaboration in the $\Upsilon$ region \cite{Shen:2013okm}. It is observed that the result of the fit with three $\rho$-like resonances 
(see previous subsection) crosses well above these data points. Since the difference with data in the $c\bar{c}$ region is not very large one could expect that, by including 
additional resonances, a reasonable good agreement may also be obtained eventually in this region. Disagreement with data in the bottomonium region is much worse and does not 
hint for such a possibility.

A natural way to approach to the infinite tower of resonance states is provided by the large-$N_C$ limit of QCD \cite{LargeNc}. Still, there is an obliged model dependence in 
the spectrum of the theory. The study of meson form factors within this limit has been undertaken, for the pion case, in Refs.~\cite{Dominguez:2001zu, Bruch:2004py}. Our 
results, eqs.~(\ref{eq:Fit1}) and (\ref{eq:Fit2}), suggest that a sensible approximation to this problem is to approach the large-$N_C$ meson masses and widths by their PDG 
values \cite{Beringer:1900zz}. However, we have no guidance on heavier states belonging to the fourth and heavier multiplets \footnote{Moreover, there can appear unphysical 
poles in a large-$N_C$ approach to the Minkowskean region \cite{Masjuan:2007ay} which prevents to relate all poles to the resonance parameters for highly excited states.}. For 
this reason we will restore for definiteness to the Veneziano model \cite{Veneziano:1968yb} for dual-$QCD_{N_C=\infty}$ which predicts a Regge trajectory for the $\rho$-like 
states where the squared masses rise linearly with the radial quantum number, $n$. String-inspired models also derive a linear relation between the mass and width of a given 
$\rho$-excitation (see \cite{Bruch:2004py} and references therein). To accommodate the known meson masses we will allow for a subleading dependence in $1/n$~\cite{Mondejar:2007dz}
\begin{equation}
 M_n^2\,=\,A\,+\,B\,n\,+\,\frac{C}{n}\,,
\end{equation}
and fit $A$, $B$ and $C$ to the PDG masses of the first three $\rho$-like states. Assuming $M_n/\Gamma_n(M_n^2)$ to be a constant, we will have at large $n$
\begin{equation}
 \Gamma_n(M_n^2)\,=\,E \sqrt{n}\,+\,F\,,
\end{equation}
with $E$ and $F$ chosen to reproduce the $\rho$ and $\rho^\prime$ on-shell widths. In this way the excited states overlap more and more with increasing 
energy, as predicted by the $N_C\to\infty$ limit of QCD. From our results in eqs.~(\ref{eq:Fit1}) and (\ref{eq:Fit2}) one may guess that $d_{s_i}$ can be assumed to be a 
constant as first approximation, while $F_{V_i}$ -which has the largest impact- decreases slightly with $i$ (perhaps alternating sign) and the opposite for $d_{M_i}$ 
($d_{m_i}$ will have no influence whatsoever). We have assessed that the emerging picture is basically independent on our assumptions on the couplings of the excited 
resonances \footnote{For instance, the best fit results and couplings do not change assuming alternating or definite sign for $F_{V_i}$, or allowing for a variation of 
$d_{M_i}$ with $i$.}. The largest impact is due to the precise value of $F_V$ that we will discuss at the end. We present our results for \footnote{$F_{V_1}$ denotes the 
coupling of the $n=2$ resonance, namely the $\rho(1450)$ meson. Analogous notation will be employed for the $\rho_i$ resonances in eq.~(\ref{eq:FF_large_Nc}).}
\begin{equation}
 F_{V_n}\,=\,(-)^n \frac{F_V}{\sqrt{n+1}}\,,
\end{equation}
assuming the remaining parameters of the higher excitations to be constant. Specifically, we consider the form factor

\begin{eqnarray}\label{eq:FF_large_Nc}
 F_V^{\omega\pi^0}(s)&=&\frac{2\sqrt{2}}{FM_VM_\omega}\left(c_{1235}m_\pi^2-c_{1256}M_\omega^2+c_{125}s\right)-\frac{4F_V}{FM_\omega}\left[d_{123}m_\pi^2+d_3(s+M_\omega^2)\right]D_\rho(s)\nonumber\\
&-&\frac{2F_{V_1}}{FM_\omega}\left(d_m m_\pi^2+d_M M_\omega^2+d_s s\right)D_{\rho^\prime}(s)-\frac{2F_{V_2}}{FM_\omega}\left(\tilde{d}_m m_\pi^2+\tilde{d}_M M_\omega^2+\tilde{d}_s s\right)D_{\rho^{\prime\prime}}(s)\nonumber\\
&-&\sum_{i=4}^{N_\infty}\frac{2F_{V_{i-1}}}{FM_\omega}\left(\tilde{d}_m m_\pi^2+\tilde{d}_M M_\omega^2+\tilde{d}_s s\right)D_{\rho_{i-1}}(s)\,,
\end{eqnarray}
where $N_\infty$ is chosen such that it includes all states with masses up the maximum energy of the considered data. With this large-$N_C$ form factor we have attempted to fit 
all available data, up to $E\sim11$ GeV. Although it can be seen in Fig.~\ref{fig:Fit_high} that the form factor (grey dotted line) can be forced to agree with the data in the 
$\Upsilon$ region, this is at the price of a worse fit to data at low and intermediate energies in Fig.~\ref{fig:Fit_low}. This suggests that the resonance approach cannot 
be naively extended into higher energy regions. On the contrary, if we only include data in the charmonium region, a good fit is possible in both energy regimes (shown as orange dashed line in 
Figs.~\ref{fig:Fit_low} and \ref{fig:Fit_high}), giving
\begin{eqnarray}\label{eq:Fit3}
 c_{1256} & = & -0.051\pm0.002\,,\quad d_3=-0.181\pm0.003\,,\quad d_M=0.82\pm0.06\,,\nonumber\\
d_s & = & -0.31\pm0.02\,, \,\quad \tilde{d}_M=\,1.53\pm0.12,\quad \tilde{d}_s=-0.31\pm0.03\,,
 \end{eqnarray}
with $\chi^2/ndf=1.1$.

We interpret this fact through the understanding that light-flavoured resonance exchanges can explain $\sigma(e^+e^-\to$hadrons) at most up to the opening of mesons made up 
of heavy quarks \footnote{Obviously, the fact there is no data in the $2-3$ GeV region restricts the determination of the dynamics associated to the fourth and heavier 
multiplets. If some data points were measured there, it would help to refine our large-$N_C$ approach.}, where new degrees of freedom that we are ignoring -and that should 
contribute to the considered process- become dynamical. In this way, our fit eq.~(\ref{eq:Fit3}) should be matched with a curve describing the charmonium data around the 
$J/\Psi$ region and extending up to the data in the $\Upsilon$ region. In this way we would have a form factor capable of describing the $\omega\pi$ cross-section from 
threshold up to $11$ GeV.

In Table \ref{Tab:1} we compute the effects on our fitted parameters produced by letting $F_V$ to vary a $10\%$ around the prediction $F_V=\sqrt{3}F$. As we have mentioned before, 
the value of $F_V$ is, by far, the most important source of uncertainty within our approach. As it can be observed, the values of resonance couplings are compatible with the 
results obtained in Refs.~\cite{Guo:2008sh, Chen:2013nna}, in particular taking into account that we have included data at higher energies provided by the SND Collaboration. 
Another interesting feature concerns the fact that, focusing in the low- and intermediate-energy data (as in \cite{Guo:2008sh, Chen:2013nna}), the fit does not change 
considerably the values of the lightest resonance couplings of the theory.

\begin{table*}[h!]
\begin{center}
\begin{tabular}{|c|c|c|c|}
\hline 
$F_V$ & $0.9F\sqrt{3}$ &  $F\sqrt{3}$&  $1.1F\sqrt{3}$\\
\hline
$c_{1256}$& $-0.045\pm0.001$ & $-0.051\pm0.002$ & $-0.056\pm0.002$\\
$d_3$ &$-0.199\pm0.003$ &$-0.181\pm0.003$ & $-0.164\pm0.003$\\
$d_M$ &$0.99\pm0.06$ & $0.82\pm0.06$ & $0.68\pm0.06$\\
$d_s$ & $-0.36\pm0.01$ &$-0.31\pm0.02$ & $-0.27\pm0.02$\\
$\tilde{d}_M$&$1.45\pm0.11$ &$1.53\pm0.12$ &$1.57\pm0.13$\\
$\tilde{d}_s$&$-0.29\pm0.03$ &$-0.31\pm0.03$ &$-0.32\pm0.03$ \\
\hline
$\chi^2/ndf$& $1.2$ & $1.1$ & $1.1$ \\
\hline
\end{tabular}
\caption{\label{Tab:1} \small{Best fit values obtained varying $F_V$ around its predicted value of $\sqrt{3}F$.}}
\end{center}
\end{table*}

\subsection{Data in quarkonium region: Matching the resonance and perturbative regimes}\label{Matching}

Although our best fit result in eq.~(\ref{eq:Fit3}) follows data closely from threshold up to $\sim3.5$ GeV, and the results at low and intermediate energies are largely 
independent on the modelization of our large-$N_C$ approach, the resonance contributions are not able to provide the suppression required by the data at higher energies. 
Therefore, we assume there is a squared energy scale, $s_0$, which splits the resonance-driven physics from the perturbative regime around which we can match both 
descriptions. However, we admit that we have not come up with a theoretical description capable of giving the observed suppression in the bottomonium region. As a 
consequence, we will consider equation (\ref{eq:FF_large_Nc}) for $s\leq s_0$ and use the simple ansatz~\footnote{This kind of power suppression is expected from the operator 
product expansion of QCD. From this point of view, $A$ will encode non-perturbative physics parametrized in terms of some hadronic matrix elements.}
\begin{equation}\label{eq:FF_asymp}
 F_V^{\omega\pi^0}(s)\,=\,\frac{A}{s^b}\,,
\end{equation}
for $s\geq s_0$, in such a way that the complex number $A$ is determined by demanding continuity for $F_V^{\omega\pi^0}(s)$, while $s_0$ and $b$ are fitted to data. We expect 
$b\sim2$, according to Belle's analysis \cite{Shen:2013okm}, $s_0\geq2$ GeV, because of our results in eq.~(\ref{eq:Fit3}) (see also Fig. \ref{fig:Fit_high}) and 
$s_0< M_{J/\Psi}\sim 3.1$ GeV since new degrees of freedom that we are ignoring become excited at these energies.

\begin{figure}[ht!]
\begin{center}
\vspace*{1.25cm}
\includegraphics[scale=0.5,angle=-90]{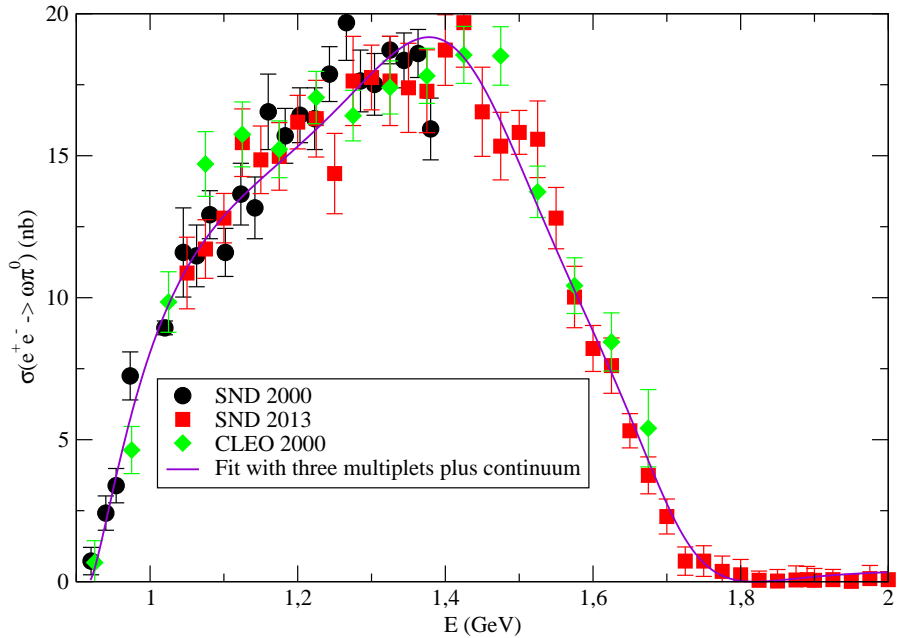}
\caption[]{\label{fig:Fit_low_FINAL} \small{SND \cite{Achasov:2000wy, Achasov:2013btb} and isospin-rotated CLEO data \cite{Edwards:1999fj} for $\sigma(e^+e^-\to \omega \pi^0)$ 
below $2$ GeV are confronted to our best fit results. The form factor includes the contribution of three $\rho$-like resonances and the continuum. The data in the quarkonium 
region, which can be seen in Fig.~\ref{fig:Fit_high_FINAL}, are taken into account in the fit}.}
\end{center}
\end{figure}

\begin{figure}[ht!]
\begin{center}
\vspace*{1.25cm}
\includegraphics[scale=0.5,angle=-90]{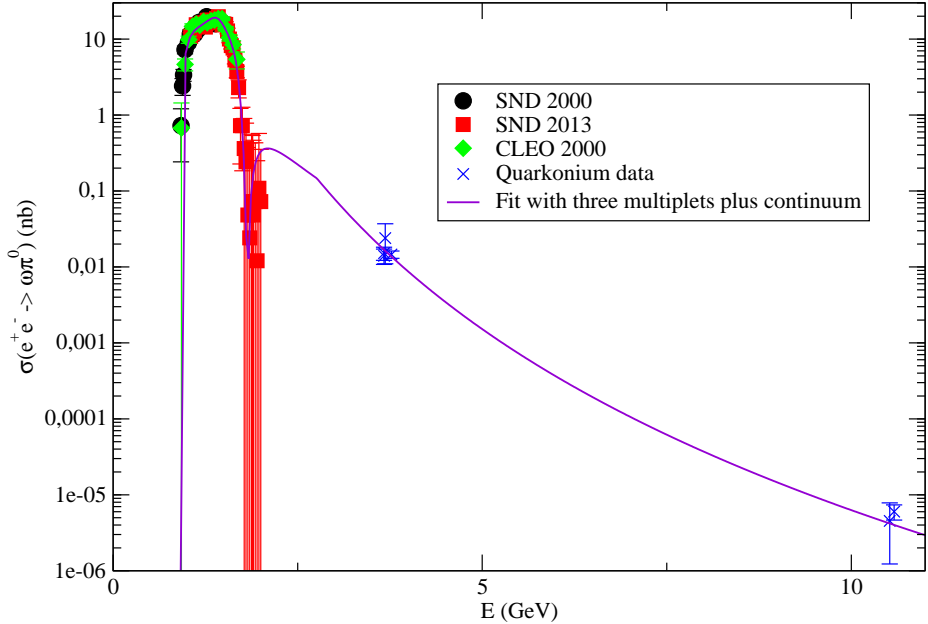}
\caption[]{\label{fig:Fit_high_FINAL} \small{SND \cite{Achasov:2000wy, Achasov:2013btb}, isospin-rotated CLEO data \cite{Edwards:1999fj}, CLEO data in the charmonium region 
\cite{Adam:2004pr, Adams:2005ks} and Belle \cite{Shen:2013okm} data in the bottomonium region for $\sigma(e^+e^-\to \omega \pi^0)$ are confronted to our best fit results. The form factor includes  
 three $\rho$-like resonances plus a continuum contribution.}}
\end{center}
\end{figure}

In order to preserve good agreement with data it is necessary to keep the contribution of the $\rho(1700)$ resonance. On the other hand, it is not clear whether we should maintain the contribution given by the sum over the tower of higher 
resonances simultaneously with the asymptotic contribution of eq.~(\ref{eq:FF_asymp}). Therefore, we present both groups of results in tables \ref{Tab:2} and \ref{Tab:3}. 
According to them, a more robust description is obtained by considering that the $\rho(770)$, $\rho(1450)$ and $\rho(1700)$ resonances essentially saturate all resonance 
contributions to the process \footnote{Measurement of data in the $\left[2,3\right]$ GeV region may demand the contribution of a fourth resonance, as it appears to be the case in the 
$e^+e^-\to\pi^+\pi^-$ BaBar analysis \cite{Lees:2012cj}.}, while the non-resonant dynamics can be parametrized well by means of a contribution to $F_V^{\omega\pi^0}(s)$ that falls as $1/s^2$, 
as pointed out in the Belle analysis \cite{Shen:2013okm}. The matching scale for both descriptions, $s_0$, is not determined with precision due to the lack of data in the 
$2-3$ GeV region but according to the orange dashed line in Fig.~\ref{fig:Fit_high}, it should lie just below the $J/\Psi$. The good agreement between 
the values shown in table \ref{Tab:3} and fit (\ref{eq:Fit2}) supports the consistency of the picture. In table \ref{Tab:4} we consider the error associated to the choice of 
$F_V$ for a $1/s^2$ damping of the asymptotic contribution. The error introduced by the variation of $b$ in eq.~(\ref{eq:FF_asymp}) around $2$ ($b=1.96\pm0.07$ in table 
\ref{Tab:3}) is negligible compared to that induced by $F_V$. We will consider the errors shown in table \ref{Tab:4} as those of our best fit form factor given by the second 
column in table \ref{Tab:3}
. Our best fit results are plotted in 
Figs.~\ref{fig:Fit_low_FINAL} and \ref{fig:Fit_high_FINAL}.

\begin{table*}[h!]
\begin{center}
\begin{tabular}{|c|c|c|c|}
\hline
$c_{1256}$& $-0.045\pm0.006$ & $-0.052\pm0.003$ & $-0.047\pm0.004$\\
$d_3$ &$-0.158\pm0.007$ &$-0.200\pm0.005$ & $-0.156\pm0.008$\\
$d_M$ &$0.40\pm0.12$ & $0.84\pm0.19$ & $0.38\pm0.14$\\
$d_s$ & $-0.15\pm0.04$ &$-0.45\pm0.05$ & $-0.14\pm0.05$\\
$\tilde{d}_M$&$1.99\pm0.26$ &$0.02\pm0.10$ &$2.15\pm0.18$\\
$\tilde{d}_s$&$-0.40\pm0.05$ &$-0.08\pm0.01$ &$-0.43\pm0.04$\\
$b$& $2.86\pm0.10$ & $2$ & $3$ \\
$\sqrt{s_0}$ (GeV) & $1.74\pm0.03$ & $1.87\pm0.03$ & $1.77\pm0.02$ \\
\hline
$\chi^2/ndf$& $1.2$ & $1.6$ & $1.2$ \\
\hline
\end{tabular}
\caption{\label{Tab:2} \small{Best fit values obtained by adding the asymptotic behaviour of eq.~(\ref{eq:FF_asymp}) to the large-$N_C$ description of eq.~(\ref{eq:FF_large_Nc}). 
The results of the three columns correspond, respectively, to the cases when $b$ is taken as a free parameter or $b=2$ or $3$  in the fit.}}
\end{center}
\end{table*}

\begin{table*}[h!]
\begin{center}
\begin{tabular}{|c|c|c|c|}
\hline
$c_{1256}$& $-0.055\pm0.004$ & $-0.055\pm0.004$\\
$d_3$ &$-0.180\pm0.005$ &$-0.180\pm0.005$\\
$d_M$ &$0.86\pm0.12$ & $0.86\pm0.13$\\
$d_s$ & $-0.33\pm0.04$ &$-0.33\pm0.04$\\
$\tilde{F}_{V_1}$& $0.11\pm0.07$ & $0.11\pm0.07$\\
$\tilde{d}_M$&$1.44\pm0.04$ &$1.50\pm0.04$\\
$\tilde{d}_s$&$-0.29\pm0.08$ &$-0.30\pm0.09$\\
$b$& $1.96\pm0.07$ & $2$\\
$\sqrt{s_0}$ (GeV) & $2.70^{+0.92}_{-0.38}$ & $2.76^{+1.12}_{-0.39}$\\
\hline
$\chi^2/ndf$& $1.2$ & $1.2$\\
\hline
\end{tabular}
\caption{\label{Tab:3} \small{Best fit values obtained adding the asymptotic behaviour of eq.~(\ref{eq:FF_asymp}) to the contribution of the first three $\rho$-like states in 
eq.~(\ref{eq:FF_large_Nc}). The results of the two columns correspond, respectively, to the cases when $b$ is taken as a free parameter or $b=2$ in the fit.}}
\end{center}
\end{table*}

\begin{table*}[h!]
\begin{center}
\begin{tabular}{|c|c|c|c|}
\hline 
$F_V$ & $0.9F\sqrt{3}$ &  $F\sqrt{3}$&  $1.1F\sqrt{3}$\\
\hline
$c_{1256}$& $-0.054\pm0.004$ & $-0.055\pm0.004$ & $-0.056\pm0.004$\\
$d_3$ &$-0.200\pm0.005$ &$-0.180\pm0.005$ & $-0.163\pm0.004$\\
$d_M$ &$0.92\pm0.12$ & $0.86\pm0.12$ & $0.80\pm0.12$\\
$d_s$ & $-0.34\pm0.03$ &$-0.33\pm0.04$ & $-0.32\pm0.03$\\
$\tilde{F}_{V_1}$& $0.10\pm0.08$ & $0.11\pm0.07$ & $0.12\pm0.06$\\
$\tilde{d}_M$&$1.55\pm0.05$ &$1.50\pm0.04$ &$1.41\pm0.04$\\
$\tilde{d}_s$&$-0.31\pm0.01$ &$-0.33\pm0.04$ &$-0.29\pm0.08$ \\
$\sqrt{s_0}$ (GeV) & $2.67^{+0.56}_{-0.31}$ & $2.76^{+1.12}_{-0.39}$ & $2.85^{+0.55}_{-0.49}$\\
\hline
$\chi^2/ndf$& $1.2$ & $1.2$ & $1.2$\\
\hline
\end{tabular}
\caption{\label{Tab:4} \small{Best fit values obtained when we allow a $10\%$ variation of $F_V$ around its predicted value $F_V=\sqrt{3}F$. We assume an asymptotic behaviour 
of $F_V^{\omega\pi^0}(s)$ given by eq.~(\ref{eq:FF_asymp}) with $b=2$ together with the contribution of the first three $\rho$-like states in eq.~(\ref{eq:FF_large_Nc}).}}
\end{center}
\end{table*}

\section{The $\pi\gamma^{(\star)}\gamma^{(\star)}$ form factor in $\RCT$}\label{piggFF}

The $\pi\gamma^{(\star)}\gamma^{(\star)}$ form factor is defined in terms of the vector-vector-pseudoscalar QCD three-point Green function \cite{Kampf:2011ty, RuizFemenia:2003hm},
\begin{equation}\label{VVP GF's}
 {\Pi_{VVP}}^{(abc)}_{\mu\nu}(p,q)\,=\,\int d^4x\int d^4y\mathrm{e}^{i(p\cdot x+q\cdot y)}\left\langle0\Big|T\left[V_\mu^a(x)V_\nu^b(y)P^c(0)\right]\Big|0\right\rangle\,,
\end{equation}
with vector and pseudoscalar currents defined as
\begin{equation}\label{currents}
 V_\mu^a(x)\,=\,\left(\bar{\Psi}\gamma_\mu\frac{\lambda^a}{2}\Psi\right)(x)\,,\quad P^a(x)\,=\,\left(\bar{\Psi i\gamma_5}\frac{\lambda^a}{2}\Psi\right)(x)\,.
\end{equation}
In the $SU(3)_V$ limit it reads
\begin{equation}\label{Pi_VVP}
 {\Pi_{VVP}}^{(abc)}_{\mu\nu}(p,q)\,=\,\varepsilon_{\mu\nu\alpha\beta}p^\alpha q^\beta d^{abc} \Pi_{VVP}(p^2,q^2,r^2)\,,
\end{equation}
where $r^\mu=-(p+q)^\mu$ has been defined as the momentum of the pseudoscalar density. In terms of this Green function, one can define the $\gamma\gamma\pi$ form factor (in 
the chiral limit)
 \begin{equation}\label{GF & FF}
 \mathcal{F}_{\pi^0\gamma\gamma}(p^2,q^2,r^2)\,=\,\frac{2}{3}\frac{r^2}{F}\frac{\Pi_{VVP}(p^2,q^2,r^2)}{B}\,.
\end{equation}

{\it A priori}, the contribution of pseudoscalar resonances to the $VVP$ Green function cannot be neglected. In fact, it has been shown \cite{Kampf:2011ty}  that their presence is necessary 
to fulfill consistently all operator product expansion constraints on this function and related form factors (see Ref.~\cite{JJYo} for more details). Consequently, we will take 
this contribution into account in what follows.

If all particles are virtual, the function $\Pi_{VVP}(p^2,q^2,r^2)$ \cite{RuizFemenia:2003hm, Kampf:2011ty} allows us to write the $\gamma^*(p)\gamma^*(q)\pi^*(r)$ form factor 
as~\footnote{We have defined $P_3\equiv d_m\kappa^{VVP}$ in eq.~(\ref{pig*g* FF virtual pion}), see eq.~(\ref{L RChT P KN}). $P_1$ and $P_2$ were defined in the discussion 
around eq.~(\ref{consistency condition for P_2}). Unlike $P_1$ and $P_2$, $P_3$ is unrestricted by high-energy relations and so it needs to be determined phenomenologically, 
as it is done in eq.~(\ref{P3}).
}

\begin{eqnarray}\label{pig*g* FF virtual pion}
 \mathcal{F}_{\pi^0\gamma\gamma}(p^2,q^2,r^2) & = & \frac{2r^2}{3F}\left[-\frac{N_C}{8\pi^2r^2}+4F_V^2\frac{d_3(p^2+q^2)}{(M_V^2-p^2)(M_V^2-q^2)r^2}+\frac{4F_V^2d_{123}}{(M_V^2-p^2)(M_V^2-q^2)}\right.\\
& & \left. -2\sqrt{2}\frac{F_V}{M_V}\frac{r^2 c_{1235}-p^2 c_{1256}+q^2 c_{125}}{(M_V^2-p^2)r^2}-2\sqrt{2}\frac{F_V}{M_V}\frac{r^2 c_{1235}-q^2c_{1256}+p^2 c_{125}}{(M_V^2-q^2)r^2}+\frac{64 P_1}{M_P^2-r^2}\right.\nonumber\\
& & \left. -\frac{16\sqrt{2}P_2 F_V}{(M_V^2-p^2)(M_P^2-r^2)}-\frac{16\sqrt{2}P_2 F_V}{(M_V^2-q^2)(M_P^2-r^2)}
 +\frac{16F_V^2P_3}{(M_V^2-p^2)(M_V^2-q^2)(M_P^2-r^2)}\right]\,,\nonumber
\end{eqnarray}
which displays the symmetry under the exchange of the photon momenta. In addition to the lightest vector resonances and pseudoscalar mesons we have also included the contribution 
from the lightest pseudoscalar resonances in eq.~(\ref{pig*g* FF virtual pion}) following Ref.~\cite{Kampf:2011ty}. The effect of excited pseudoscalar and vector resonances 
has been neglected. It is straightforward to check that our fully off-shell form factor, eq.~(\ref{pig*g* FF virtual pion}), is identical to the Kampf and Novotny's form factor, after 
using eqs.~(35) and (43) in Ref.~\cite{Kampf:2011ty} and eq.~(3) in Ref.~\cite{JJYo}.


Assuming the pion to be on-shell yields (we are working in the chiral limit)
\begin{eqnarray}\label{pig*g* FF real pion}
 \mathcal{F}_{\pi^0\gamma\gamma}(p^2,q^2,0) & = & \frac{2}{3F}\left[-\frac{N_C}{8\pi^2}+\frac{4F_V^2d_3(p^2+q^2)}{(M_V^2-p^2)(M_V^2-q^2)}+2\sqrt{2}\frac{F_V}{M_V}\frac{p^2c_{1256}-q^2c_{125}}{M_V^2-p^2}\right.\nonumber\\
& & \left. 
+2\sqrt{2}\frac{F_V}{M_V}\frac{q^2c_{1256}-p^2c_{125}}{M_V^2-q^2}\right]\,.
\end{eqnarray}
We note that, if the on-shell condition for the pion is assumed, the form factor in eq.~(\ref{pig*g* FF real pion}) depends on the couplings $F_V$, $c_{125}$, $c_{1256}$ and $d_3$, 
all of them constrained by the short-distance QCD information. The main differences between our analysis of the $\pi$TFF and the one reported in 
Ref.~\cite{Kampf:2011ty} are first that in this reference the short-distance constraint $F_V=\sqrt{3}F$ was not realized and, instead, the phenomenological value 
$F_V\,=\,F_\rho\,=\,(146.3\pm1.2)$ MeV was used and second, that we will include Belle's data on the $\pi$TFF, which was published after Ref.~\cite{Kampf:2011ty} was released. We note 
however that the violation of the high-energy restriction for $F_V$ is $\sim8.4\%$ within its $10\%$ range of variation that we have been using to estimate 
the main error of our approach coming from the precise value of $F_V$. We therefore expect to agree reasonably with Ref.~\cite{Kampf:2011ty} in our fit of this form 
factor in section \ref{Analysis pig*g} and on its effect in $a_\mu^{\pi^0,\,HLbL}$ in section \ref{LbL}.

The form factor with a virtual pion depends, additionally, on the five coupling combinations: $c_{1235}$, $d_{123}$, $P_1$ and $P_2$, which are restricted by asymptotic QCD 
constraints, (\ref{SD}) and (\ref{consistency condition for P_2}), and $P_3$ which shall be fixed phenomenologically. 
Indeed, the combined analyses of the $\pi(1300)\to\gamma\gamma$ and $\pi(1300)\to\rho\gamma$ decays in Ref.~\cite{Kampf:2011ty} allows to fix $\kappa^{VVP}$. Following this 
procedure, we obtain
\begin{equation}\label{P3}
 P_3\,=\,\left(-1.2\pm0.3\right)\cdot10^{-2}\, \mathrm{GeV}^2\,.
\end{equation}


From this discussion, a controlled uncertainty in the form factor in eq.~(\ref{pig*g* FF virtual pion}) can be expected, since all but one of the participating couplings are 
predicted from short-distance QCD constraints and the other one is determined within a $25\%$ accuracy. This will translate to our evaluation of the pion exchange contribution to 
the anomalous magnetic moment of the muon in section \ref{LbL}. It will be interesting to use both form factors, eqs. (\ref{pig*g* FF virtual pion}) and 
(\ref{pig*g* FF real pion}) in order to estimate the error associated to assuming a real pion exchanged in the loop, i.e. to obtaining the pion pole contribution as an 
approximation to the whole pion exchange contribution \cite{Jegerlehner:2007xe, Nyffeler:2009tw}. 

\section{Analysis of data on the $\pi\gamma^\star\gamma$ form factor}\label{Analysis pig*g}

The $\pi$TFF has been measured by the CELLO \cite{Behrend:1990sr}, CLEO \cite{Gronberg:1997fj}, BaBar \cite{Aubert:2009mc} and Belle \cite{Uehara:2012ag} 
collaborations in $e^+e^-\to e^+e^-\pi^0$, where the $\pi^0$ is produced by two photons exchanged in $t$-channel~\cite{Brodsky:1971ud}. Only one of the final-state leptons was tagged and the other 
one escaped the detector in a small-angle emission, ensuring a high virtuality for one of the photons and almost on-shellness for the other. Therefore, this is considered to 
be a measurement of the the $\pi\gamma^\star\gamma$ form factor. Since the momenta of both photons are Euclidean in those experiments, we should replace $q^2\to -Q^2$, 
$p^2\to-P^2$ in eq.~(\ref{pig*g* FF real pion}) to obtain the form factors with an on-shell pion, and assume $P^2=0$ as an accurate approximation to the experimental 
detection conditions. These data, particularly the BaBar and Belle measurements with photon virtualities up to $Q^2\sim40$ GeV$^2$ and thus probing the (pre-)asymptotic limit 
of QCD, have triggered a lot of attention recently and a number of analyses using various approaches \cite{piggTFF}.

Once the short-distance QCD constraints for $F_V$, $c_{125}$, $c_{1256}$ (\ref{SD}) and $d_3$ (\ref{equivalent_SD_d3}) are implemented into eq.~(\ref{pig*g* FF real pion}), the 
form factor can be conveniently rewritten as
\begin{equation}\label{fit piggFF}
 \mathcal{F}_{\pi^0\gamma\gamma}(Q^2)\,=\,-\frac{F}{3}\frac{Q^2\left(1+32\sqrt{2}\frac{P_2F_V}{F^2}\right)+\frac{N_C}{4\pi^2}\frac{M_V^4}{F^2}}{M_V^2(M_V^2+Q^2)}\,,
\end{equation}
in agreement with Ref.~\cite{Kampf:2011ty}. It should be pointed out that the dependence on the pseudoscalar resonance coupling $P_2$ in eq.~(\ref{fit piggFF}) is introduced 
through the use of eq.~(\ref{equivalent_SD_d3}) because the $\pi$TFF  depends only upon vector resonance couplings and $F$ (for the pseudo-Goldstone dynamics).

We observe, however, that sticking to the Brodsky-Lepage constraint for $P_2$, eq.~(\ref{consistency condition for P_2}), does not yield a satisfactory description of the 
data. We find, in accord with Ref.~\cite{Kampf:2011ty}, that a small violation of this equation, $\sim4\%$, yields the best fit to the $\pi$TFF data. 
Specifically, we obtain
\begin{equation}\label{P2}
P_2\,=\,-\left(1.13\pm0.12\right)\cdot 10^{-3}\,\mathrm{GeV}\,,\quad \chi^2/dof=1.01\,,
\end{equation}
where the error is dominated by the $10\%$ variation of $F_V$ around its predicted value of $\sqrt{3}F$ (\ref{SD}) ($c_{1256}$ and $d_3$ change according to its value). This 
result is compatible with the value in Ref.~\cite{Kampf:2011ty} \footnote{It must be noted that Belle data \cite{Uehara:2012ag}, which seems to agree better with the Brodsky-Lepage 
asymptotic prediction than BaBar data \cite{Aubert:2009mc}, was not available when Ref.~\cite{Kampf:2011ty} was published.}
\begin{equation}
P_2\,=\,-\left(1.21\pm0.03\right)\cdot 10^{-3}\,\mathrm{GeV}\,,
\end{equation}
where the errors are those stemming from the minimization procedure only.

We point out that fits of similar quality could be obtained by neglecting pseudoscalar resonance effects and considering the first excited vector multiplet instead. In 
particular, fitting only $\widetilde{c_{1256}}$ (defined in analogy to the coupling $c_{1256}$ for the first multiplet) yields $\widetilde{c_{1256}}=-\left(1.75\pm0.01\right)\cdot10^{-3}$ 
with again a $\chi^2$ per degree of freedom of order unity. This ambiguity may explain why one can find in the literature approaches where pseudoscalar resonances are ignored 
and two vector multiplets are considered instead or settings where only the first multiplet of pseudoscalar and vector resonances is accounted for. It does not seem possible 
to settle this issue soon, even with more precise data on the $\pi$TFF. Our argument to prefer the description including the pseudoscalar mesons and only the lightest 
multiplet of pseudoscalar and vector resonances is the consistency of short-distance constraints in the odd-intrinsic parity resonance chiral Lagrangian that can be achieved 
in the single resonance approximation~\cite{JJYo}.

In any case, more accurate measurements of this form factor at large momentum transfer are needed to elucidate whether the Brodsky-Lepage-like asymptotic behaviour (approached 
by Belle) or its violation (hinted by BaBar) describe the high-energy data. In the next section we will propose an ideally-suited observable to probe 
$\mathcal{F}_{\pi^0\gamma\gamma}$ with both photons off their mass-shell.

\begin{figure}[h!]
\begin{center}
\vspace*{1.25cm}
\includegraphics[scale=0.5,angle=-90]{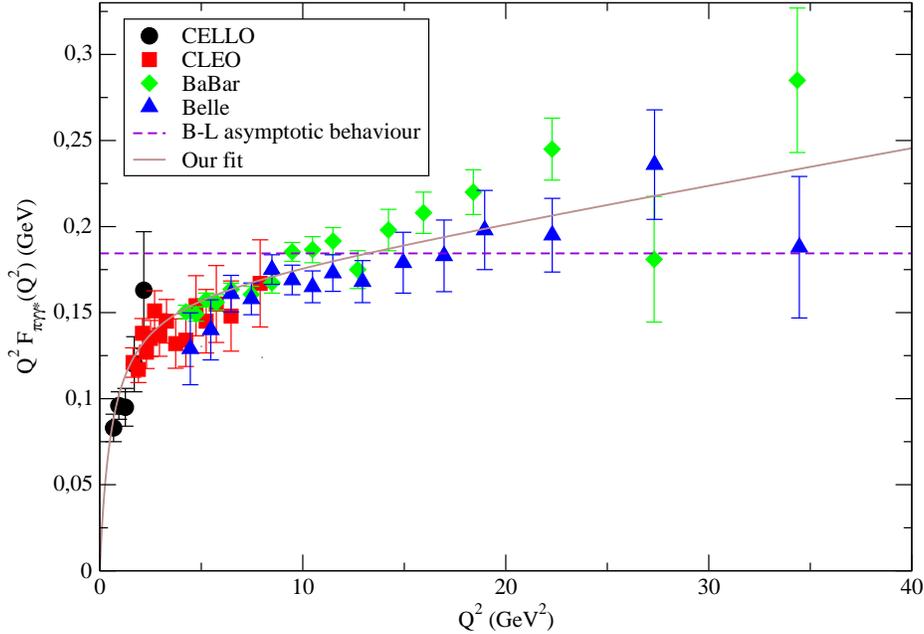}
\caption[]{\label{fig:piggFF} \small{CELLO \cite{Behrend:1990sr}, CLEO \cite{Gronberg:1997fj}, BaBar \cite{Aubert:2009mc} and Belle \cite{Uehara:2012ag} data for the $\pi$TFF 
are confronted to our best fit result using the form factor in eq.~(\ref{pig*g* FF real pion}) as explained in the main text. The error band associated to the $10\%$ variation 
of $F_V$ cannot be appreciated.}}
\end{center}
\end{figure}

\section{A genuine probe of the $\pi\gamma^\star\gamma^\star$ form factor}\label{probe of pig*g*}

In the previous sections we have seen that the $\gamma\gamma^* P$ and $\gamma^\star \omega\pi^0$ form factors require the contributions of one and three multiplet of vector 
resonances, respectively, to account for experimental data. The reason of this behavior is that the specific on-shell particles involved in the process determine the possible 
$VVP$ couplings that are necessary. In this section we shall study some processes involving the pseudoscalar TFF with two virtual photons which, eventually, may provide 
information on the couplings of excited vector resonances.

The $\pi\gamma^\star\gamma^\star$ form factor can be probed in the process $e^+(q_+)e^-(q_-)\to\gamma^\star(k)\to\pi^0(p_\pi)\gamma^\star(k^\prime)\to\pi^0(p_\pi)\mu^+(p_ +)\mu^-(p_-)$ 
which, to our knowledge, has not been studied or searched for previously\footnote{This process occurs only via the $s$-channel. A similar contribution to $e^+e^- \to \pi^0e^+e^-$ 
is suppressed by experimental kinematical considerations.}. This decay can be measured by the KLOE Collaboration for $k^2, k'^2\lesssim 1$ GeV$^2$ and in Belle-II for photon 
virtualities up to some ($10.5$ GeV)$^2$. In this process both photons are time-like as opposed to the $t$-channel extraction of the $\pi$TFF discussed in Sects.~\ref{piggFF} 
and \ref{Analysis pig*g}. It should also be noted that the form factor that takes part in the evaluation of the $a_\mu^{\pi^0,\,HLbL}$ also has both time-like photons. The 
additional uncertainty induced by the non-vanishing $\Gamma_\rho(s)$ for $s>4m_\pi^2$ should not be in principle a limitation to probe the $\pi$TFF studying 
$e^+e^-\to\pi^0\mu^+\mu^-$. At the present level of precision for $\pi$TFF, any possible quark-hadron duality violation \cite{Shifman:2000jv} in relating the Euclidean and 
Minkowskean regions shall be neglected.

In terms of suitable invariants \cite{Kumar:1970cr},
\begin{equation}\label{invariants}
 s\equiv k^2\,,\quad s_1\equiv{k^\prime}^2\,,\quad t_0\equiv (q_+-p_\pi)^2\,,\quad t_1\equiv (q_+-p_+)^2\,,\quad u_1\equiv(k-p_+)^2\,,
\end{equation}
the corresponding spin-averaged and unpolarized squared matrix element reads
\begin{eqnarray}\label{m.e. summed and averaged invariants}
  \sum \overline{\Big|\mathcal{M}\Big|^2} & = & \frac{512\alpha^4\pi^4}{s^2 s_1^2}\Big|\mathcal{F}_{\pi^0\gamma\gamma}(k^2,{k^\prime}^2)\Big|^2\left\lbrace -2 m_\mu^4 s^2+
m_\mu^2 s \left[m_\mu^4+m_\mu^2 \left(m_\pi^2+s+s_1-2t_0-4t_1+2u_1\right)\right.\right.\nonumber\\
& & \left. \left. +m_\pi^4 +m_\pi^2 \left(-3 s+s_1-3 t_0-2 t_1+ u_1\right)+3 s^2-4 s s_1+5 s t_0+6 s t_1- 3 s u_1+s_1^2-3 s_1 t_0\right.\right.\nonumber\\
& & \left. \left. -2 s_1 t_1+ s_1 u_1+3 t_0^2+4 t_0 t_1-2 t_0 u_1+4 t_1^2-4 t_1 u_1 + u_1^2\right]+\frac{1}{4}\left[2 s \left(s_1-2 m_\mu^2\right) (s+t_1-u_1) \right.\right.\nonumber\\
& & \left. \left. \left(-m_\mu^2-m_\pi^2+s+t_0+t_1\right)+4 \left(m_\mu^2-t_1\right) (s+t_1-u_1) \left(m_\mu^2+m_\pi^2-s-t_0-t_1\right)\right.\right.\nonumber\\
& & \left. \left. (s-s_1+t_0+t_1-u_1)+s \left(s_1-2 m_\mu^2\right) \left(-m_\mu^2-m_\pi^2+s+t_0+t_1\right)^2-2 (s+t_1-u_1)^2 \right.\right.\nonumber\\
& & \left. \left. \left(-m_\mu^2-m_\pi^2+s+t_0+t_1\right)^2-2 s^2 \left(s_1-2 m_\mu^2\right)^2-2 \left(m_\mu^2-t_1\right)^2 (s-s_1+t_0+t_1-u_1)^2\right.\right.\nonumber\\
& & \left. \left. +s \left(s_1-2 m_\mu^2\right) (s-s_1+t_0+t_1-u_1)^2+2 s \left(2 m_\mu^2-s_1\right) \left(m_\mu^2-t_1\right) (s-s_1+t_0+t_1-u_1)\right.\right.\nonumber\\
& & \left. \left. +s \left(s_1-2 m_\mu^2\right) (s+t_1-u_1)^2+s \left(s_1-2 m_\mu^2\right) \left(m_\mu^2-t_1\right)^2\right]\right\rbrace \,,
\end{eqnarray}
where we have neglected the electron mass. Since the flavour facilities can measure this cross-section at very small values of $k^2$ --close to the threshold of 
$(2m_\mu+m_\pi)^2$-- we kept $m_\mu\neq0$ and $m_\pi\neq0$ in eq.~(\ref{m.e. summed and averaged invariants}) as we have done in the numerics.
The cross-section can be written \cite{Kumar:1970cr}
\begin{equation}
 \sigma=\frac{1}{2^7\pi^4s^2}\int_{4m_\mu^2}^{(\sqrt{s}-m_\pi)^2}\frac{\mathrm{d}s_1}{\lambda^{1/2}(s,s_1,m_\pi^2)}\int_{t_0^-}^{t_0^+}\frac{\mathrm{d}t_0}{\sqrt{1-\xi^2}}\int_{u_1^-}^{u_1^+}\frac{\mathrm{d}u_1}{\lambda^{1/2}(s,m_\mu^2,u_1)\sqrt{1-\eta^2}}\int_{t_1^-}^{t_1^+}\frac{\mathrm{d}t_1\overline{\Big|\mathcal{M}\Big|^2}}{\sqrt{1-\zeta^2}}\,,
\end{equation}
with the definitions
\begin{eqnarray}
 \zeta & =& (\omega-\xi\eta)\left[(1-\xi^2)(1-\eta^2)\right]^{-1/2}\,,\quad \omega=(s-m_\mu^2-u_1+2t_1)\lambda^{-1/2}(s,m_\mu^2,u_1)\,,\\
 \eta & =& \left[2s s_1-(s+m_\mu^2-u_1)(s+s_1-m_\pi^2)\right]\lambda^{-1/2}(s,m_\mu^2,u_1)\lambda^{-1/2}(s,s_1,m_\pi^2)\,,\quad \xi=\frac{s-m_\pi^2-s_1+2t_0}{\lambda^{1/2}(s,s_1,m_\pi^2)}\,,\nonumber
\end{eqnarray}
and the $t_0$, $u_1$ and $t_1$ integration limits
\begin{eqnarray}
 t_0^{\pm}& = & m_\pi^2-\frac{s+m_\pi^2-s_1}{2}\pm\frac{\lambda^{1/2}(s,m_\pi^2,s_1)}{2}\,,\; u_1^{\pm}=s+m_\mu^2-\frac{s+s_1-m_\pi^2}{2}\pm\frac{\sqrt{s_1(s_1-4m_\mu^2)\lambda(s,s_1,m_\pi^2)}}{2s_1}\nonumber\\
 t_1^{\pm}& = & m_\mu^2-\frac{s+m_\mu^2-u_1}{2}+\frac{\lambda^{1/2}(s,m_\mu^2,u_1)}{2}\left[\xi\eta\pm\sqrt{(1-\xi^2)(1-\eta^2)}\right]\,.
 \end{eqnarray}
Measurements of the differential cross section $d\sigma/ds_1$ for different values of $s$ can be used to measure $\pi^0\gamma^*\gamma^*$ the full form factor in a clean way.

\begin{figure}[h!]
\begin{center}
\vspace*{1.25cm}
\includegraphics[scale=0.5,angle=-90]{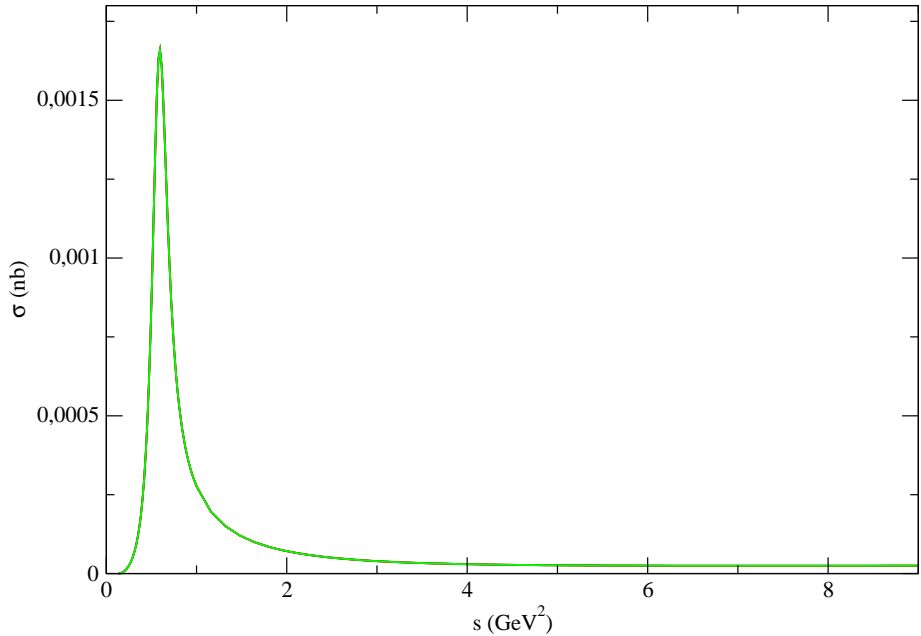}
\caption[]{\label{fig:sigmas} \small{Our predictions for $\sigma(e^+e^-\to\pi^0\mu^+\mu^-)(s)$ are plotted using the values of the couplings in 
eq.~(\ref{Couplings piggFF real pion}). The very small error band cannot be appreciated.}}
\end{center}
\end{figure}

The cross-section for $e^+e^- \to \pi^0\mu^+\mu^-$ can be predicted using the form factor in eq.~(\ref{pig*g* FF real pion}) with $p^2\to s$, $q^2\to s_1$ and $(M_R^2-x)^{-1}\to D_R(x)$ 
(see eq.~(\ref{D_R(x)})) by employing the values of the couplings discussed in the previous section:
\begin{eqnarray}\label{Couplings piggFF real pion}
 F_V & = & \sqrt{3}F(1.0\pm0.1)\,,\quad c_{125}\,=\,0\,,\quad c_{1256}\,=\,-\frac{N_C M_V}{32\sqrt{2}\pi^2F_V}\,,\nonumber\\
 d_3 & = & -\frac{N_C M_V^2}{64\pi^2F_V^2}\,+\,\frac{F^2}{8F_V^2}\,+\,\frac{4\sqrt{2}P_2}{F_V}\,,\quad P_2 \, = \, -\left(1.13\pm0.12\right)\cdot 10^{-3}\,\mathrm{GeV}\,.
\end{eqnarray}
The central curve and the corresponding error bands (almost indistinguishable) are plotted in Fig.~\ref{fig:sigmas}. The $\rho(770)$ peak shows neatly and, at higher energies, 
the cross-section seems to approach a plateau. The possible contribution of the $\rho(1450)$ resonance (and higher excitations) and its associated 
uncertainties are negligible with the linear scales used in the plots of this section.

\begin{figure}[h!]
\begin{center}
\vspace*{1.25cm}
\includegraphics[scale=0.5,angle=-90]{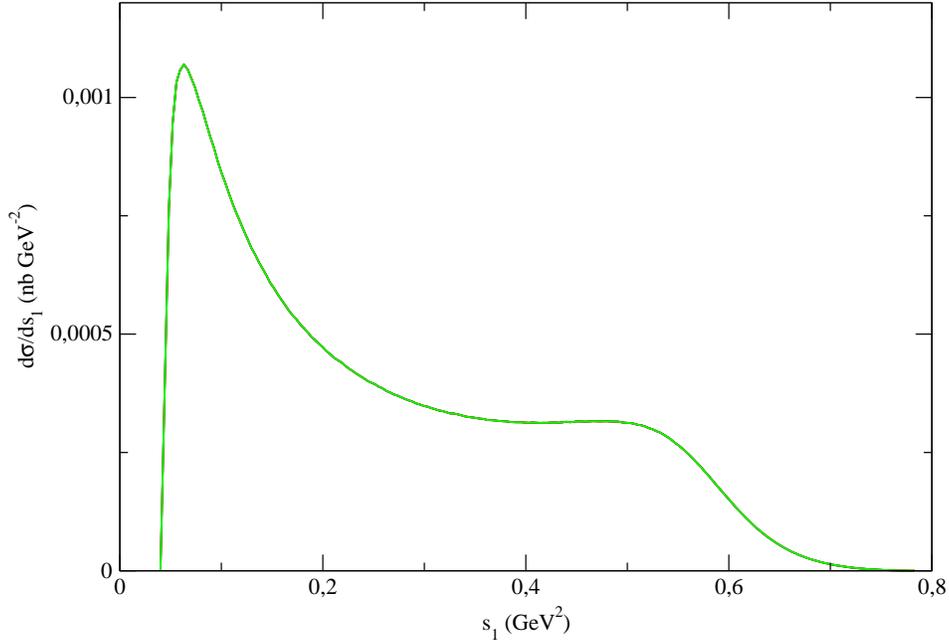}
\caption[]{\label{fig:sigmas1} \small{Our predictions for $\mu^+\mu^-$ distribution at $s=(1.02$ GeV$)^2$ are plotted using the values of the couplings in eq.~(\ref{Couplings 
piggFF real pion}). The error band cannot be appreciated.}}
\end{center}
\end{figure}

As it was pointed out previously, the differential cross-section as a function of the muon pair invariant mass can be measured at different values of the center-of-mass energy, 
$s$. The characteristic shape of this distribution is shown in Figure \ref{fig:sigmas}. This profile makes its measurement at KLOE-2 specially appealing and, for this reason, 
it is plotted for $s=M_\phi^2$ in Fig.~\ref{fig:sigmas1}. The analogous plot at $s=M_{\Upsilon(4S)}^2$, corresponding to B-factories, is not shown. However, it will be very valuable to measure some points at high virtualities in the muon pair 
invariant mass distribution to check the predicted asymptotic behaviours.

The proposed observables of the $e^+e^-\to\pi^0\mu^+\mu^-$ process can provide complementary information on the $\pi$TFF data. Measurements of the $\mu^+\mu^-$ invariant 
mass distribution at KLOE-2 and Belle-II and new, more precise data at high $Q^2$ on $\sigma(e^+e^-\to\pi^0e^+e^-)$ would be most beneficial in improving our understanding of 
the pion exchange contribution to $a_\mu^{HLbL}$, which is evaluated in the next section according to our findings in sections \ref{omegapigammaFF} to \ref{Analysis pig*g}.

\begin{figure}[h!]
\begin{center}
\vspace*{1.25cm}
\includegraphics[scale=0.5,angle=-90]{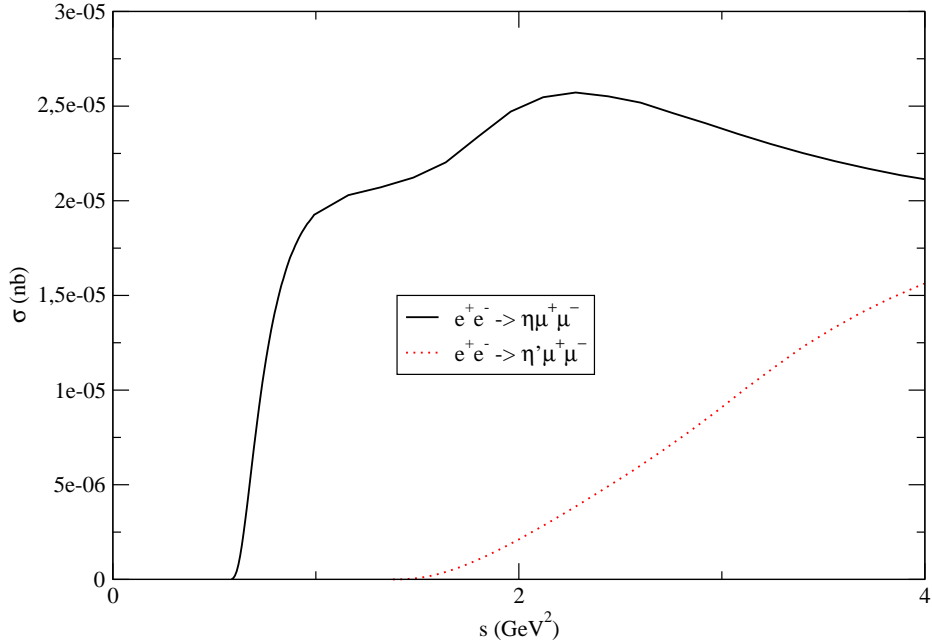}
\caption[]{\label{fig:sigmas_eta} \small{Our predictions for $e^+e^-\to\eta^{(\prime)}\mu^+\mu^-$ cross-section are plotted using the values of the couplings in 
eq.~(\ref{Couplings piggFF real pion}) and eqs.~(\ref{definitions_Cq_Cs})-(\ref{relation_etaTFF_piTFF}) for the $\eta$-$\eta^\prime$ mixing.}}
\end{center}
\end{figure}

\begin{figure}[h!]
\begin{center}
\vspace*{1.25cm}
\includegraphics[scale=0.5,angle=-90]{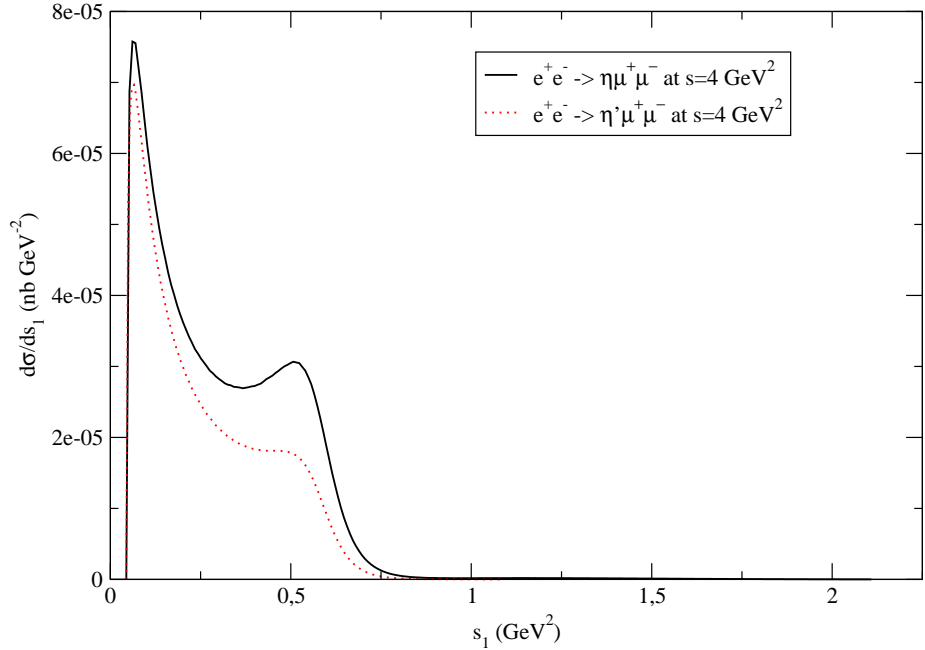}
\caption[]{\label{fig:sigmas1_eta} \small{Our predictions for the $\mu^+\mu^-$ distribution in the $e^+e^-\to\eta^{(\prime)}\mu^+\mu^-$ processes are plotted at $s=4$ GeV$^2$ 
using the values of the couplings in eq.~(\ref{Couplings piggFF real pion}) and eqs.~(\ref{definitions_Cq_Cs})-(\ref{relation_etaTFF_piTFF}) for the $\eta$-$\eta^\prime$ mixing.}}
\end{center}
\end{figure}

Similar processes with $\eta$, $\eta'$ replacing the $\pi^0$ meson production could in principle provide measurements of the $\gamma^*\gamma^*\eta^{(')}$ form factors. The 
relationship between the $(\eta,\eta')$TFF and the $\pi$TFF is given in Section \ref{etaetap}.  The results for the total and differential cross sections shown in Figures \ref{fig:sigmas_eta} and \ref{fig:sigmas1_eta} 
are obtained using eqs.~(\ref{definitions_Cq_Cs})-(\ref{relation_etaTFF_piTFF}) and the $\pi$TFF discussed in the paragraph above eq.~(\ref{Couplings piggFF real pion}). 
The effect of the contribution of higher excited states is negligible in the d$\sigma$/d$s_1$ distributions and is at the same level induced by the uncertainties on the 
$\eta$-$\eta^\prime$ mixing in the cross-section plot. They are of order  $30(20)\%$ for the $\eta(\eta^\prime)$ cases. With respect to the observable considered in Figure 
\ref{fig:sigmas1_eta}, we point out that at KLOE-2 ($s=M_\phi^2=1.04$ GeV$^2$) the $\eta$ distribution will be less prominent and no hadronic structure will show up because 
there is not enough phase space available, while the corresponding process for the $\eta^\prime$ could not even be produced at these energies. The $\mu^+\mu^-$ distribution 
at $s=4$ GeV$^2$ that we present in Figure \ref{fig:sigmas1_eta} shows a characteristic structure produced by the $\rho(770)$ meson contribution and can nevertheless be 
measured either using energy-scan at the Novosibirsk CMD and SND experiments or using the radiative return method \cite{ISR} at B-factories, like Belle-II.

\section{Consequences for the pion exchange contribution to the hadronic light-by-light muon $g-2$}\label{LbL}

For more than a decade, the anomalous magnetic moment of the muon $a_{\mu}$ has shown a persistent discrepancy between the BNL measurement \cite{BNL} and the theoretical 
predictions \cite{Miller:2007kk, Jegerlehner:2009ry, Hagiwara:2011af} (both of them have a similar uncertainty of $\sim6.3\cdot10^{-10}$) at the three sigma level. The 
Standard Theory value of $a_\mu$ receives contributions from QED, electroweak and QCD processes. Although the first one accounts for most of the anomaly \cite{Aoyama:2012wk}, 
the theoretical uncertainty is completely dominated by the hadronic contributions. The latter is essentially saturated by the hadronic vacuum polarization at LO, which at 
present can be better obtained via $e^+e^-$ hadroproduction or hadronic tau decays (via isospin rotation \cite{Alemany:1997tn, Davier:2002dy, Davier:2009ag}; see also 
\cite{Jegerlehner:2011ti, Benayoun:2012wc, Blum:2013xva}). The hadronic light-by-light HLbL contribution, although smaller, contributes to $a_{\mu}$ with a similar uncertainty 
as the LO hadronic vacuum polarization. While the error bar in the latter would in principle be reduced with more accurate measurements of the hadronic cross-section, the 
second one is fully theoretical, coming from the various models used to evaluate this contribution \cite{Knecht:2001qf, HLbL} (see Ref.~\cite{Czyz:2013sga} for an updated 
report on this problem). The need to reduce the uncertainty of the hadronic contribution to $a_\mu$, particularly the one due to HLbL, is increased in view of the upcoming 
experiments at Fermilab and J-Parc that expect to improve the current accuracy by a factor of four \cite{Talks}, down to $1.6\cdot10^{-10}$, clearly smaller than the error of 
the Standard Theory determination.

The HLbL scattering contribution to $a_\mu$ involves the $\left\langle VVVV \right\rangle$ Green function connected to three off-shell photons \cite{Jegerlehner:2009ry, 
Nyffeler:2009tw}. The non-trivial interplay of different regions of momenta leads to a mixing of long- and short-distance contributions in which its splitting in parts to be 
computed in terms of quarks and hadrons, respectively, is cumbersome and avoiding double-counting becomes a problematic issue. A classification of the different contributions 
relying on the chiral and large-$N_C$ countings was put forward in Ref.~\cite{deRafael:1993za}. According to it, the dressed \footnote{In general, all interactions of hadrons 
and quarks with photons are dressed by form factors, e.g. via $\rho-\gamma$ mixing.} charged pion loop is leading in the chiral counting but subleading in the $1/N_C$-expansion. 
At NLO in the chiral expansion, but leading in $1/N_C$ there appear the pseudoscalar meson exchanges dominated by the $\pi^0$ contribution. Also leading in 
$1/N_C$ but next-to-next-to-leading in the chiral expansion there are contributions from other resonances ($f_0$, $a_1$, ...) and from the dressed quark loop~\cite{deRafael:1993za}. 
Although the separation of the different contributions is ambiguous and model-dependent, there is consensus in the literature that the pseudoscalar exchange contributions (and 
in particular, that of the $\pi^0$) give the most of the $a_\mu^{HLbL}$ value, a feature which is not understood on the basis of the combined chiral and $1/N_C$-counting  
introduced above.

We will evaluate this dominant $a_\mu^{\pi^0,HLbL}$ contribution employing the $\pi$TFF derived in section \ref{piggFF}. For this, the fully off-shell form 
factor in eq.~(\ref{pig*g* FF virtual pion}) is needed. To illustrate the error due to assuming a real pion (this corresponds to pinning down the pion pole contribution 
from the whole pion exchange contribution) we will also employ the corresponding form factor in eq.~(\ref{pig*g* FF real pion}), with an on-shell pion \footnote{Melnikov and 
Vainsthein (in \cite{HLbL}) pointed out that this procedure violates momentum conservation at the external vertex and propose to use the constant form factor derived from the Wess-Zumino-Witten 
action to obtain consistently the pion-pole contribution to $a_\mu^{\pi^0,HLbL}$. Since many references in the literature use momentum-dependent form factors to obtain the 
on-shell pion-pole contribution (thus violating momentum conservation at the external photon vertex), we have used this approach to illustrate the effect of the associated 
error.}. The main formulae needed for this evaluation (two- and three-dimensional integrations, respectively) are given in the Appendix. The form factor with a real pion will 
be fixed using eqs.~(\ref{Couplings piggFF real pion}). For the fully off-shell form factor, short-distance constraints (\ref{SD}) will be employed to determine $c_{1235}$ 
and $d_{123}$ as well as $P_1=0$, which is also required by consistency with QCD asymptotics. Finally $P_3$ will be set to eq.~(\ref{P3}). The error will be estimated by the 
quoted variations of $F_V$ and $P_2$, using $F=(92.20\pm0.14)$ MeV~\cite{Beringer:1900zz} and by the uncertainty on the value of the form factor at the origin, discussed 
below eq.~(\ref{Result virtual pion}).

In this way, using the (incorrect momentum-dependent) form factor for the external vertex  we obtain
\begin{equation}\label{Result real pion}
 a_{\mu}^{\pi^0,HLbL}\,=\,\left(5.75\pm0.06\right)\cdot 10^{-10}\,,
\end{equation}
for the pion pole contribution, and
\begin{equation}\label{Result virtual pion}
 a_{\mu}^{\pi^0,HLbL}\,=\,\left(6.66\pm0.21\right)\cdot 10^{-10}
\end{equation}
for the whole pion exchange contribution, which implies that putting the pion on-shell underestimates the value of $a_{\mu}^{\pi^0,HLbL}$ by $\sim14\%$ and the the corresponding error by a 
factor of four (similar numbers are obtained using other approaches). Contrary to what happens in all observables that we have considered, the error of our evaluation of 
$a_{\mu}^{\pi^0,HLbL}$ in eq.~(\ref{Result real pion}) is not dominated by the value of $F_V$ (the error induced by $P_2$ is also negligible). The uncertainty quoted in 
eq.~(\ref{Result real pion}) is essentially given by a contribution encoding the very low-energy Physics: the chiral corrections to $\pi$TFF at the origin. We have evaluated 
the latter using \cite{RuizFemenia:2003hm}
\begin{equation}\label{Correction to the FF at the origin}
 \mathcal{F}_{\pi\gamma\gamma}(0)\,=\,-\frac{N_C}{4\pi^2F}(1-\Delta)\,,
\end{equation}
with
\begin{equation}\label{Delta}
 \Delta\,=\,\frac{4\pi^2}{3}\frac{F^2}{M_V^2}\frac{m_\pi^2}{M_V^2}\sim 5.9\cdot 10^{-3}\,,
\end{equation}
where the short-distance QCD constraints for $c_{1235}$ and $d_{123}$ in eq.~(\ref{SD}) were used. This value of $\Delta$ implies a shift in $a_{\mu}^{\pi^0,HLbL}$ of $-0.07\cdot 10^{-10}$. 
For this reason, the central value of eq.~(\ref{Result real pion}) has been allocated in the center of the error band. Corrections to eq.~(\ref{Delta}) should be suppressed by 
further powers of $m_\pi^2/M_V^2$ and shall be neglected. If, instead of relying on the $R\chi T$ prediction, eqs.~(\ref{Correction to the FF at the origin}) and (\ref{Delta}), 
we restored to the measured value of $\Gamma(\pi^0\to\gamma\gamma)$, the bound on $\Delta$ would be a factor of five (three) larger according to the PDG \cite{Beringer:1900zz} 
(the PrimEx experiment~\cite{Larin:2010kq}). Its forthcoming measurement at KLOE-2 \cite{Babusci:2011bg} should provide soon a determination capable of testing eq.~(\ref{Delta}) 
and, therefore, of reducing the uncertainty on the determination of $a_{\mu}^{\pi^0,HLbL}$ within a given approach. The error of another low-energy quantity, $F$, has a much 
smaller influence on the error in eq.~(\ref{Result real pion}): $\pm0.02\cdot 10^{-10}$.

The uncertainty quoted in eq.~(\ref{Result virtual pion}) for the virtual pion case, on the contrary, receives three comparable contributions: from 
$\left\lbrace F_V,\, P_2\right\rbrace$, from $P_3$ and from $\mathcal{F}_{\pi\gamma\gamma}(0)$ (the effect of the error of $F$ is $\sim1/3$ with respect to the others and the 
influence of the precise value of the pseudoscalar resonance mass is marginal). Since the error of $\left\lbrace F_V,\, P_2\right\rbrace$ is determined by the range allowed for 
$F_V$, more precise phenomenological analyses may help to reduce this uncertainty. The incertitude on $P_3$ is given by the limit $BR(\pi^\prime\to\gamma\gamma)<72$ eV, set by 
Belle~\cite{Belle_piprime}. A more stringent bound on this decay width will also help to reduce the error of eq.~(\ref{Result virtual pion}). The prospects for reducing the 
error on $\mathcal{F}_{\pi\gamma\gamma}(0)$ were already discussed in the previous paragraph.

Our result, eq.~(\ref{Result virtual pion}), is compared to other determinations in Table \ref{Other_results}, where the method employed in each of them is also given for 
reference.

\begin{table*}[h!]
\begin{center}
\begin{tabular}{|c|c|c|c|}
\hline 
$a_\mu^{\pi^0,HLbL}\cdot 10^{10}$ & Method and Reference\\
\hline
$5.58\pm0.05$ & \small{Extended NJL Model \cite{NJL} (Bijnens, Pallante and Prades in \cite{HLbL})}\\
$5.56\pm0.01$ & \small{Naive VMD Model (Hayakawa, Kinoshita [and Sanda] in \cite{HLbL})}\\
$5.8\pm1.0$ & \small{Large-$N_C$ with two vector multiplets, $\pi$-pole contribution~\cite{Knecht:2001qf}}\\
$7.7\pm1.0$ &\small{Large-$N_C$ with two vector multiplets, $\pi$-pole contribution (Melnikov and Vainshtein in \cite{HLbL})}\\
$7.2\pm1.2$ & \small{$\pi$-exchange contribution corresponding to~\cite{Knecht:2001qf} evaluated in~\cite{Nyffeler:2009tw} (Jegerlehner and Nyffeler)}\\
$6.9$ & \small{Holographic models of QCD~\cite{Hong:2009zw}}\\
$6.54\pm0.25$ & \small{Holographic models of QCD~\cite{Cappiello:2010uy}}\\
$6.58\pm0.12$ & \small{Lightest Pseudoscalar and Vector Resonance saturation~\cite{Kampf:2011ty}}\\
$6.49\pm0.56$ & \small{Rational Approximants \cite{Masjuan:2012qn}}\\
$5.0\pm0.4$ & \small{Non-local chiral quark model~\cite{Dorokhov:2012qa}}\\
\hline
$6.66\pm0.21$ & \small{This work, short-distance constraints of \cite{Kampf:2011ty} revisited and data set updated}\\
\hline
\end{tabular}
\caption{\label{Other_results} \small{Our result for $a_{\mu}^{\pi^0,HLbL}$ in eq.~(\ref{Result virtual pion}) is compared to other determinations. 
The method employed in each of them is also given. We specify those works that approximate $a_\mu^{\pi^0,HLbL}$ by the pion pole contribution. It is understood that all 
others consider the complete pion exchange contribution.}}
\end{center}
\end{table*}

\section{$\eta$ and $\eta^\prime$ exchange contributions to the hadronic light-by-light muon $g-2$}\label{etaetap}

In this section we evaluate the contributions of the next lightest pseudoscalar mesons ($\eta$ and $\eta^\prime$) to $a_{\mu}^{HLbL}$. In order to do that we need to relate 
the respective TFF to the $\pi$TFF. We will treat the $\eta$-$\eta^\prime$ mixing in the two-angle mixing scheme (consistent with the large-$N_C$ limit of QCD~\cite{KL}) and 
work in the quark flavour basis~\cite{twoanglemixing} where
\begin{equation}\label{diag_u}
 \mathrm{diag}(u)\,=\,\left( \frac{\pi^0+C_q\eta+C_{q^\prime}\eta^\prime}{\sqrt{2}},\; \frac{-\pi^0+C_q\eta+C_{q^\prime}\eta^\prime}{\sqrt{2}},\; -C_s\eta+C_{s^\prime}\eta^\prime\right)\,,
\end{equation}
in which
\begin{eqnarray}\label{definitions_Cq_Cs}
 C_q & \equiv & \frac{F}{\sqrt{3}\mathrm{cos}(\theta_8-\theta_0)}\left(\frac{\mathrm{cos}\theta_0}{f_8}-\frac{\sqrt{2}\mathrm{sin}\theta_8}{f_0}\right)\,,\quad C_{q^\prime}\equiv\frac{F}{\sqrt{3}\mathrm{cos}(\theta_8-\theta_0)}\left(\frac{\sqrt{2}\mathrm{cos}\theta_8}{f_0}+\frac{\mathrm{sin}\theta_0}{f_8}\right)\,,\nonumber\\
 C_s & \equiv & \frac{F}{\sqrt{3}\mathrm{cos}(\theta_8-\theta_0)}\left(\frac{\sqrt{2}\mathrm{cos}\theta_0}{f_8}+\frac{\mathrm{sin}\theta_8}{f_0}\right)\,,\quad C_{s^\prime}\equiv\frac{F}{\sqrt{3}\mathrm{cos}(\theta_8-\theta_0)}\left(\frac{\mathrm{cos}\theta_8}{f_0}-\frac{\sqrt{2}\mathrm{sin}\theta_0}{f_8}\right)\,.\nonumber\\
\end{eqnarray}
The values of the pairs of decay constants and mixing angles are~\cite{twoanglemixing} 
\begin{equation}\label{values_mixingangless&decayscts}
 \theta_8\,=\,\left(-21.2\pm1.6\right)^\circ,\quad  \theta_0\,=\,\left(-9.2\pm1.7\right)^\circ,\quad f_8\,=\, \left(1.26\pm0.04\right)F,\quad f_0\,=\, \left(1.17\pm0.03\right)F\,.
\end{equation}
We will consider these errors as independent in the following.  

Within this mixing scheme, the $\eta$ and $\eta^\prime$ TFF can be easily related to the $\pi$TFF
\begin{eqnarray}\label{relation_etaTFF_piTFF}
 \mathcal{F}_{\eta\gamma\gamma}(p^2,\,q^2,\,r^2) & = & \left(\frac{5}{3}C_q-\frac{\sqrt{2}}{3}C_s\right)\mathcal{F}_{\pi\gamma\gamma}(p^2,\,q^2,\,r^2)\,,\nonumber\\
 \mathcal{F}_{\eta^\prime\gamma\gamma}(p^2,\,q^2,\,r^2) & = & \left(\frac{5}{3}C_{q^\prime}+\frac{\sqrt{2}}{3}C_{s^\prime}\right)\mathcal{F}_{\pi\gamma\gamma}(p^2,\,q^2,\,r^2)\,.
\end{eqnarray}

\begin{figure}[h!]
\begin{center}
\vspace*{1.25cm}
\includegraphics[scale=0.5,angle=-90]{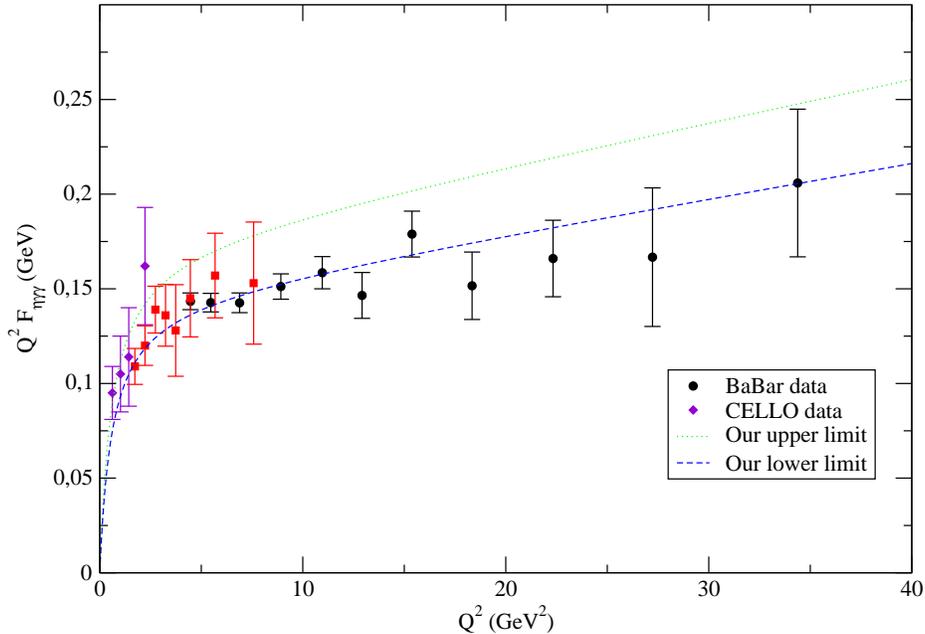}
\caption[]{\label{fig:etaTFF} \small{Our predictions for the $\eta$TFF using the values of the couplings in eq.~(\ref{Couplings piggFF real pion}) and the $\eta$-$\eta^\prime$ 
mixing in eq.~(\ref{values_mixingangless&decayscts}) are confronted to BaBar~\cite{BABAR:2011ad}, CELLO~\cite{Behrend:1990sr} and CLEO~\cite{Gronberg:1997fj} data. The error 
band is completely dominated by the uncertainty on the $\eta$-$\eta^\prime$ mixing.}}
\end{center}
\end{figure}

\begin{figure}[h!]
\begin{center}
\vspace*{1.25cm}
\includegraphics[scale=0.5,angle=-90]{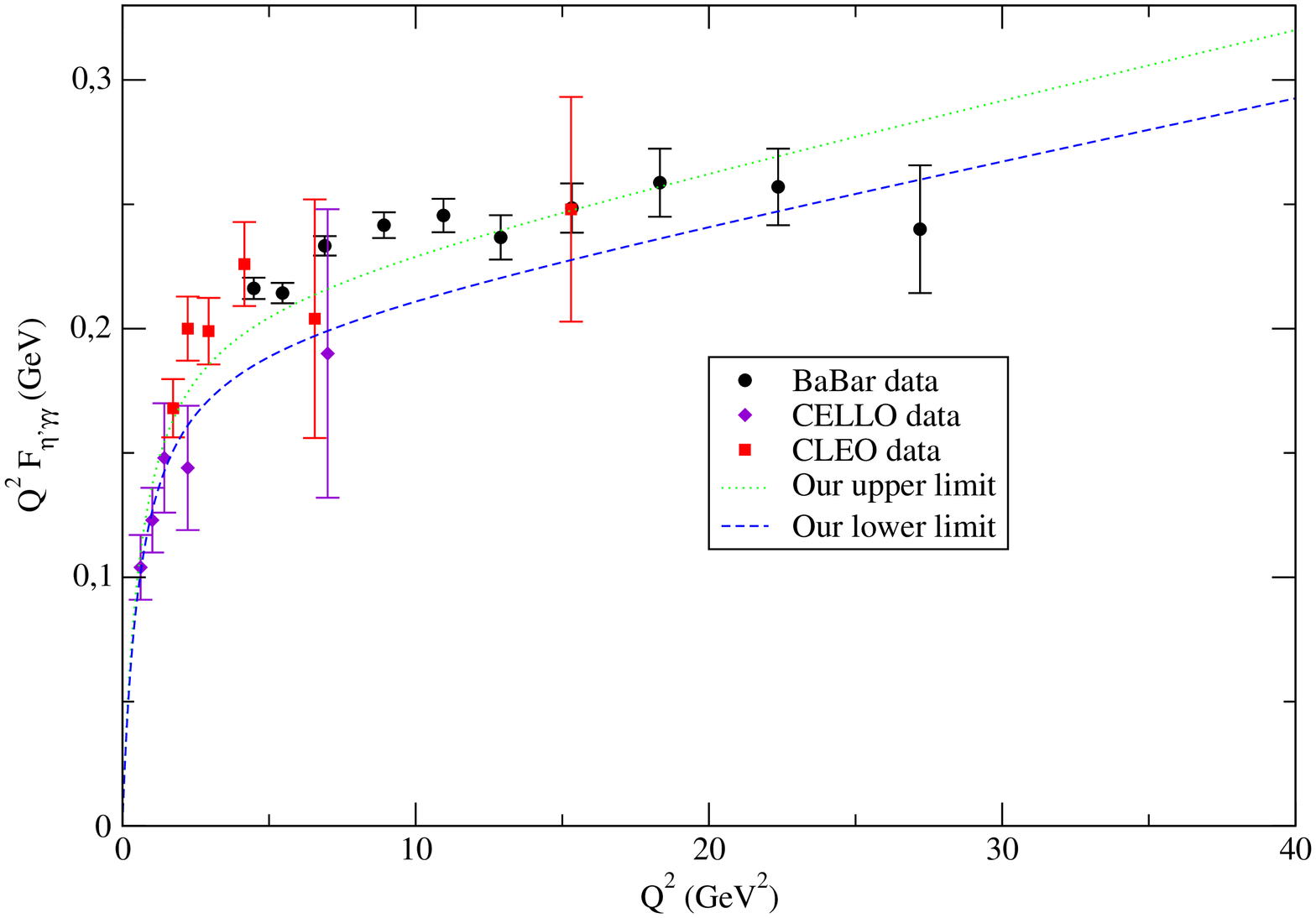}
\caption[]{\label{fig:etapTFF} \small{Our predictions for the $\eta^\prime$TFF using the values of the couplings in eq.~(\ref{Couplings piggFF real pion}) and the $\eta$-$\eta^\prime$ 
mixing in eq.~(\ref{values_mixingangless&decayscts}) are confronted to BaBar~\cite{BABAR:2011ad}, CELLO~\cite{Behrend:1990sr} and CLEO~\cite{Gronberg:1997fj} data. The error 
band is completely dominated by the uncertainty on the $\eta$-$\eta^\prime$ mixing.}}
\end{center}
\end{figure}

We have therefore predicted the $\eta$ and $\eta^\prime$ TFF using our results for the $\pi$TFF. 
The corresponding error is completely dominated by the $\eta$-$\eta^\prime$ 
mixing. In Figs.~\ref{fig:etaTFF} and \ref{fig:etapTFF} we confront them to BaBar~\cite{BABAR:2011ad}, CELLO~\cite{Behrend:1990sr} and CLEO~\cite{Gronberg:1997fj} data. In the 
case of the $\eta$TFF good agreement can be seen, although BaBar data tend to lie in the border of our predicted lower limit. Even though data from different experiments on 
the $\eta^\prime$TFF show slight tension, the overall agreement of our prediction with them is quite good. We observe that our $\pi$TFF-based prediction tends to show a tiny 
larger slope than the $\eta$ and $\eta^\prime$ TFF data. This feature may be caused by BaBar data on the $\pi$TFF. It remains to be seen if new, more accurate, measurements 
of these TFF confirm this tendency or not. As a rule of thumb, the comparison of our result for $a_{\mu}^{\pi^0,HLbL}$ (both with BaBar and Belle data on the $\pi$TFF) with 
the one in Ref.~\cite{Kampf:2011ty} (only with BaBar data) suggests that this effect is accounted for in the quoted error.

Then we have evaluated the $\eta$ and $\eta^\prime$ pole and pion exchange contributions to $a_{\mu}^{HLbL}$ as explained in section~\ref{LbL} with the results

\begin{equation}\label{Result real eta}
 a_{\mu}^{\eta,HLbL}\,=\,\left(1.44\pm0.26\right)\cdot 10^{-10}\,,\quad a_{\mu}^{\eta^\prime,HLbL}\,=\,\left(1.08\pm0.09\right)\cdot 10^{-10}
\end{equation}
for the pole contribution and
\begin{equation}\label{Result virtual eta}
 a_{\mu}^{\eta,HLbL}\,=\,\left(2.04\pm0.44\right)\cdot 10^{-10}\,,\quad a_{\mu}^{\eta^\prime,HLbL}\,=\,\left(1.77\pm0.23\right)\cdot 10^{-10}
\end{equation}
for the whole exchange contribution. As it happened in the $\pi^0$ case, the $\eta^{(\prime)}$-pole approximation underestimates clearly the $HLbL$ contribution, by $\sim 30(45)\%$, 
and the error, by a factor of roughly two. This is confirmed by comparing our results in eq.~(\ref{Result real eta}) with those obtained in Ref.~\cite{Escribano:2013kba}
\begin{equation}\label{Result real eta Pere}
 a_{\mu}^{\eta,HLbL}\,=\,\left(1.38\pm0.16\right)\cdot 10^{-10}\,,\quad a_{\mu}^{\eta^\prime,HLbL}\,=\,\left(1.22\pm0.09\right)\cdot 10^{-10}\,
\end{equation}
which agree within errors.

Taking into account our determinations of $a_{\mu}^{\pi^0,HLbL}$ (\ref{Result virtual pion}), $a_{\mu}^{\eta,HLbL}$ and $a_{\mu}^{\eta^\prime,HLbL}$(\ref{Result virtual eta}), 
we obtain for the contribution of the three lightest pseudoscalars
\begin{equation}\label{Result virtual P}
 a_{\mu}^{P,HLbL}\,=\,\left(10.47\pm0.54\right)\cdot 10^{-10}\,.
\end{equation}
This number is compared to other determinations in the literature in Table \ref{Other_results_2}. Again, the method employed in each determination is also given for reference. 

\begin{table*}[h!]
\begin{center}
\begin{tabular}{|c|c|c|c|}
\hline 
$a_\mu^{P,HLbL}\cdot 10^{10}$ & Method and Reference\\
\hline
$8.5\pm1.3$ & \small{Extended NJL Model \cite{NJL} (Bijnens, Pallante and Prades in \cite{HLbL})}\\
$8.27\pm0.64$ & \small{Naive VMD Model (Hayakawa, Kinoshita [and Sanda] in \cite{HLbL})}\\
$8.3\pm1.2$ & \small{Large-$N_C$ with two vector multiplets, $P$-pole contribution~\cite{Knecht:2001qf}}\\
$11.4\pm1.0$ &\small{Large-$N_C$ with two vector multiplets, $P$-pole contribution (Melnikov and Vainshtein in \cite{HLbL})}\\
$9.9\pm1.6$ & \small{$\pi$-exchange contribution corresponding to~\cite{Knecht:2001qf} evaluated in~\cite{Nyffeler:2009tw} (Jegerlehner and Nyffeler)}\\
$10.7$ & \small{Holographic models of QCD~\cite{Hong:2009zw}}\\
$9.0\pm0.7$ & \small{Rational Approximants \cite{Escribano:2013kba} using half-width rule \cite{Masjuan:2012sk}, $P$-pole contribution}\\
$5.85\pm0.87$ & \small{Non-local chiral quark model~\cite{Dorokhov:2012qa}}\\
$11.4\pm1.3$ & \small{Average of various approaches (Prades, de Rafael and Vainshtein in~\cite{HLbL}}\\
\hline
$10.47\pm0.54$ & \small{This work, lightest Pseudoscalar and Vector Resonance saturation}\\
\hline
\end{tabular}
\caption{\label{Other_results_2} \small{Our result for $a_{\mu}^{P,HLbL}$ in eq.~(\ref{Result virtual P}) is compared to other determinations. 
The method employed in each of them is also given. We specify those works that approximate $a_\mu^{P,HLbL}$ by the pseudoscalar pole contribution. It is understood that all 
others consider the complete pseudoscalar exchange contribution.}}
\end{center}
\end{table*}
\section{Conclusions and outlook}\label{Conclusions}

In the framework of the $R\chi T$, we have studied the $\gamma^*\omega\pi$ and $\gamma^*\gamma^*\pi$ form factors which have some common free parameters arising from the $VV'P$ Green function in the resonance 
region. When compared to experimental data, these two form factors can provide complementary and useful information to predict 
the HLbL contribution to the anomalous magnetic moment of the muon, $a_{\mu}$.

We have first considered the energy region below 2 GeV for the $\gamma^*\omega\pi$ form factor and have found that, in agreement with  Ref.~\cite{Achasov:2013btb}, the 
inclusion of three $\rho$-like resonances is sufficient to describe well the experimental data.  We have also analyzed whether the $R\chi T$ form factor could be extended to 
higher energies and explain all existing data. It was found that, within the large-$N_C$ approach, the resonance contributions can describe well the data up to the $J/\psi$ 
region but it fails to account for the data beyond this energy scale without assuming further degrees of freedom. Data on this form factor in the bottomonium region 
\cite{Shen:2013okm} falls much faster than the $1/s$ asymptotic behavior expected in QCD. In comparison with a similar approach that uses $\tau$ decay data \cite{Guo:2008sh, Chen:2013nna}, 
our present study has improved the understanding of the $\left[1.5,2\right]$ GeV energy region thanks to the use of the $e^+e^-$ data and the implementation of two 
short-distance constraints (one of them we derived here for the first time) which were missed in the quoted references.

In the second part of the paper we have studied the $\pi$TFF within $R\chi T$ in close analogy with Ref.~\cite{Kampf:2011ty}. Our improvement with respect to this previous analysis is 
two-folded: on the one hand we have included a high-energy constraint which was not realized in that reference and, on the other, we have used Belle data , which appeared after Ref.~\cite{Kampf:2011ty} was published. Our error estimate is also more robust since, in addition to the errors on the resonance couplings, we have also included the (dominant) uncertainty introduced 
by the value of the $\pi$TFF at the origin. We have shown that it is possible to describe the $\pi$TFF data adding to the pseudo-Goldstone bosons only the lightest multiplet of 
pseudoscalar and vector resonances with tiny violations of the asymptotic constraints. We have proposed that a check of this $\pi$TFF can be done through observables 
associated to $e^+e^-\to \mu^+\mu^-\pi^0$ and we have discussed the feasibility of their measurements at KLOE-2 and Belle-II. The two photons involved in this process are 
timelike, therefore it can provide an alternative measurement of the $\gamma^*\gamma^*\pi$ vertex to the one done by the `traditional' t-channel dominant contribution to 
$e^+e^-\to e^+e^-\pi^0$ \cite{Brodsky:1971ud}. 

Finally, we have applied our results to compute the pseudoscalar exchange contribution to $a_{\mu}^{HLbL}$. Our result, 
$a_{\mu}^{\pi^0,HLbL}\,=\,\left(6.66\pm0.21\right)\cdot 10^{-10}$, is compatible with that in Ref.~\cite{Kampf:2011ty} but has a larger error as a result of including the 
uncertainty on the value of the $\pi$TFF at the origin. We have also recalled that approximating the contribution of pion exchange by that of the pion pole underestimates 
$a_{\mu}^{\pi^0,HLbL}$ by $\left[15\sim 20\right]\%$, which artificially increases the discrepancy with the BNL measurements of $a_{\mu}$ (the corresponding error is also 
undervalued). Then, using our study of the $\pi$TFF, we have predicted the $\eta$ and $\eta^\prime$ transition form factors on the basis of the $\eta-\eta'$ mixing scheme 
in the quark flavor basis. Our predictions obtained for $a_{\mu}^{\eta^{(\prime)},HLbL}$ are
\begin{eqnarray}
a_{\mu}^{\eta,HLbL}\,&=&\,\left(2.04\pm0.44\right)\cdot 10^{-10}\ , \nonumber \\ 
a_{\mu}^{\eta^\prime,HLbL}\, &=&\,\left(1.77\pm0.23\right)\cdot 10^{-10}\ .\nonumber 
\end{eqnarray}
In these cases, it is also shown that approximating the pseudoscalar exchange by the pseudoscalar pole 
contribution clearly underestimates the results and their associated errors. 

As the main result of our analysis we find that the contribution of the three lightest pseudoscalar 
mesons ($\pi^0$, $\eta$ and $\eta^\prime$) to the muon anomaly is 
\begin{equation}
a_{\mu}^{P,HLbL}\,=\,\left(10.47\pm0.54\right)\cdot 10^{-10}\ , 
\end{equation} 
in good agreement with the two reference values: $\left(9.9\pm1.6\right)\cdot 10^{-10}$ (Jegerlehner and Nyffeler \cite{Nyffeler:2009tw}) and 
$\left(11.4\pm1.3\right)\cdot 10^{-10}$ (Prades, de Rafael and Vainshtein in \cite{HLbL}). The smaller error bar of our result would decrease the uncertainty in the prediction 
of $a_{\mu}^{HLbL}$, and sligthly increase the muon $g-2$ discrepancy. If the results for the $\pi$, $K$ loops and from the contribution of scalar and axial-vector resonances
~\footnote{The heavy-quark loop contribution is taken from Jegerlehner and Nyffeler's evaluation (which coincides with the Bijnens, Pallante and Prades value in \cite{HLbL}), 
because the Prades, de Rafael and Vainshtein number only accounts for the $c$-quark loop.} are added to our result we find
\begin{equation}\label{estimation a_mu^LbL}
 a_{\mu}^{HLbL}\,=\,\left(11.8\pm2.0\right)\cdot10^{-10}\,,
\end{equation}
which basically coincides with the Jegerlehner and Nyffeler's central value,
 $a_{\mu}^{HLbL}\,=\,\left(11.6\pm4.0\right)\cdot10^{-10}$ \cite{Nyffeler:2009tw}), and with the result of Prades, de Rafael and Vainshtein, 
$ a_{\mu}^{HLbL}\,=\,\left(10.5\pm2.6\right)\cdot10^{-10}$ in \cite{HLbL}. There is a good agreement within errors with both of them. The current theoretical uncertainty of 
$a_\mu$, $\pm6.2\cdot10^{-10}$, has two dominant sources: the one coming from the hadronic vacuum polarization contribution at LO, $\pm4.7\cdot10^{-10}$, and the one induced 
by computations of the HLbL scattering contribution, $\pm4.0\cdot10^{-10}$ \cite{Nyffeler:2009tw}. If, instead of the latter, our error estimate for the HLbL scattering 
contribution is used the total uncertainty would be $\pm5.1\cdot10^{-10}$ with the central value remaining basically the same.

While lattice is progressing towards a reliable evaluation of $a_{\mu}^{HLbL}$~\cite{Blum:2013qu} only a close collaboration between theory and experiment can lead to a 
reduction of the current error on this quantity. On the theory side, a deeper study of short-distance relations derived from perturbative QCD can be helpful for this 
purpose. In particular, the study of the $\left\langle VVVV\right\rangle$ Green function in the resonance region may clarify if the asymptotic constraints demanded to the 
$\pi$TFF are complete or not~\footnote{There are also some issues concerning the evaluation of the other contributions to $a_{\mu}^{HLbL}$: the dressed pseudo-Goldstone and 
quark loop contribution. In the first one, much larger values are found (up to a factor of four) in Refs.~\cite{RM}. Four to five times larger contribution for the latter is 
obtained in Refs.~\cite{GFW, GdR}.}. Also the (subdominant) contribution of scalar and axial-vector resonances needs further studies, since its relative error is still quite 
large (see however Ref.~\cite{Pauk:2014rta}).

On the experimental side, the error of the dominant pseudoscalar exchange contribution can be reduced by more precise measurements of hadronic processes at $s\lesssim4$ GeV$^2$: 
the pseudoscalar ($\pi^0$, $\eta$ and $\eta^\prime$) TFF, the two-photon pseudoscalar decay widths and the $e^+e^-\to \mu^+\mu^- \pi^0$ observables that we have proposed in 
this work. Though indirectly, a more accurate determination of the $\eta$-$\eta^\prime$ mixing can also allow to reduce the uncertainty on the corresponding contributions to 
the muon anomaly through their relation with the more precise $\pi$TFF measurements. More accurate data on the $e^+e^-\to V\pi^0$ processes and on the $\pi$TFF at high energies 
may shed some light on the fulfillment of the asymptotic QCD predictions in hadronic processes. Among all these, the earliest improvement can be expected from the KLOE-2 measurement 
of $\pi^0\to\gamma\gamma$ \cite{Babusci:2011bg}, which should be capable of reducing the error associated to the value of the $\pi$TFF at the origin by a factor of four, at 
the $1\cdot10^{-11}$ level.

\section*{Acknowledgments}

This work has been partially funded by Conacyt and DGAPA. The support of project PAPIIT IN106913 is also acknowledged. P.~R.~ has benefited from discussions with J.~J.~Sanz 
Cillero on short-distance constraints from $\left\langle VVP\right\rangle$ and the $\pi$TFF and with P.~Masjuan on the different sources of error in the determination of the 
pseudoscalar exchange contribution to $a_{\mu}^{HLbL}$. We thank Zhi-Hui Guo and Leonid Kardapoltsev for providing us with CLEO data on the vector spectral function of the 
$\tau^-\to\omega\pi^-\nu_\tau$ decays and conversations on this topic.

\section*{Appendix}\label{appendix}

This appendix collects some formulae used for the evaluation of the pion pole/exchange contribution to the hadronic light-by-light muon anomalous magnetic moment in Sect.~\ref{LbL}. We will follow 
the notation of Ref.~\cite{Knecht:2001qf}, where angular integrations of the relevant two-loop integrals were first performed analytically using the method of Gegenbauer 
polynomials. The remaining two-dimensional integrations can be readily performed numerically provided the $\pi$TFF can be written
\begin{equation}\label{decomposition KN}
 \mathcal{F}_{\pi^0\gamma\gamma}(q_1^2,q_2^2)\,=\,\frac{F}{3}\left[f(q_1^2)-\sum_{M_{V_i}}\frac{1}{q_2^2-M_{V_i}^2}g_{M_{V_i}}(q_1^2)\right]\,.
\end{equation}
Then, the hadronic light-by-light contribution to $a_\mu$ reads
\begin{equation}\label{on-shell pi LbL 1}
 a_\mu^{\pi^0,\,HLbL}\,=\,\left(\frac{\alpha}{\pi}\right)^3\left[a_\mu^{\pi^0(1),\,HLbL}+a_\mu^{\pi^0(2),\,HLbL}\right]\,,
\end{equation}
with
\begin{equation}\label{on-shell pi LbL 2}
 a_\mu^{\pi^0(1),\,HLbL}\,=\,\int_0^\infty dQ_1 \int_0^\infty dQ_2 \left[w_{f_1}(Q_1,Q_2)\,f^{(1)}(Q_1^2,Q_2^2)+\sum_{M_{V_i}}w_{g_1}(M_{V_i},Q_1,Q_2)\,g^{(1)}_{M_{V_i}}(Q_1^2,Q_2^2)\right]\,,
\end{equation}
and
\begin{equation}\label{on-shell pi LbL 3}
 a_\mu^{LbL,\pi^0(2),\,HLbL}\,=\,\int_0^\infty dQ_1 \int_0^\infty dQ_2 \sum_{M=m_\pi,M_{V_i}}w_{g_2}(M,Q_1,Q_2)\,g^{(2)}_M(Q_1^2,Q_2^2)\,.
\end{equation}
In the previous equation, $w_{{\left\lbrace f/g\right\rbrace}_i}(q_1^2,q_2^2)$ are weight factors, whose expressions can be found in Ref.~\cite{Knecht:2001qf}. 
${\left\lbrace f/g\right\rbrace}^{(i)}$ are generalized form factors given by
\begin{eqnarray}\label{GFFs}
& & f^{(1)}(Q_1^2,Q_2^2)\,=\,\frac{F}{3}f(-Q_1^2)\,\mathcal{F}_{\pi^0\gamma\gamma}(-Q_2^2,0)\,,\quad g^{(1)}_{M_{V_i}}(Q_1^2,Q_2^2)\,=\,\frac{F}{3}\frac{g_{M_{V_i}}(-Q_1^2)}{M_{V_i}^2}\,\mathcal{F}_{\pi^0\gamma\gamma}(-Q_2^2,0)\,,\nonumber\\
& & g^{(2)}_{m_\pi}(Q_1^2,Q_2^2)\,=\,\frac{F}{3}\,\mathcal{F}_{\pi^0\gamma\gamma}(-Q_1^2,-Q_2^2)\left[f(0)+\sum_{M_{V_i}}\frac{g_{M_{V_i}(0)}}{M_{V_i}^2-m_\pi^2}\right]\,,\nonumber\\
& & g^{(2)}_{M_{V_i}}(Q_1^2,Q_2^2)\,=\,\frac{F}{3}\,\mathcal{F}_{\pi^0\gamma\gamma}(-Q_1^2,-Q_2^2)\frac{g_{M_{V_i}(0)}}{m_\pi^2-M_{V_i}^2}\,.
\end{eqnarray}
Our expressions for the $\pi$TFF in the case of virtual (\ref{pig*g* FF virtual pion}) and real pion (\ref{pig*g* FF real pion}) can indeed be written 
according to eq.~(\ref{decomposition KN})
:
\begin{eqnarray}\label{KN on-shell}
 f(q^2) & = & \frac{2}{F^2}\left[\frac{-2\sqrt{2}c_{1256}F_V(M_V^2-2q^2)}{M_V(M_V^2-q^2)}-\frac{N_C}{8\pi^2}-\frac{4d_3F_V^2}{M_V^2-q^2}
\right]\,,\nonumber\\
 g_{M_V}(q^2) & = & \frac{2}{F^2}\left[2\sqrt{2}c_{1256}F_VM_V+4d_3F_V^2\frac{M_V^2+q^2}{M_V^2-q^2}
\right]\,,
\end{eqnarray}
for on-shell pion, and the additional contributions
\begin{eqnarray}\label{KN off-shell}
 \Delta f(q^2,r^2) & = & \frac{2r^2}{F^2}
\frac{-16\sqrt{2}P_2F_V}{(M_V^2-q^2)(M_P^2-r^2)}
\,,\\
 \Delta g_{M_V}(q^2,r^2) & = & \frac{2r^2}{F^2}\left\lbrace
 \frac{4 d_{123}F_V^2}{M_V^2-q^2}-\frac{16\sqrt{2}P_2 F_V}{M_P^2-r^2}+\frac{16F_V^2P_3}{(M_V^2-q^2)(M_P^2-r^2)}\right\rbrace\nonumber
\end{eqnarray}
for the general situation in which the pion is off its mass-shell. The predicted vanishing of the $c_{1235}$, $c_{125}$ and $P_1$ couplings according to asymptotic constraints 
has already been taken into account to simplify eqs.~(\ref{KN on-shell}) and (\ref{KN off-shell}).

In the latter case, eqs.~(\ref{on-shell pi LbL 1})-(\ref{on-shell pi LbL 3}) should be replaced by \cite{Jegerlehner:2009ry}
\begin{eqnarray}\label{off-shell pi LbL 1}
  a_\mu^{\pi^0,\,HLbL} & = & -\frac{2\alpha^3}{3\pi^2}\int_0^\infty dQ_1 \int_0^\infty dQ_2\int_{-1}^{+1} dt \sqrt{1-t^2} Q_1^3 Q_2^3 \left[\frac{F_1(Q_1^2,Q_2^2,t)}{Q_2^2+m_\pi^2}I_1(Q_1,Q_2,t)\right.\nonumber\\
& & \left. +\frac{F_2(Q_1^2,Q_2^2,t)}{Q_3^2+m_\pi^2}I_2(Q_1,Q_2,t)\right]\,,
\end{eqnarray}
where $Q_3=(Q_1+Q_2)$, $t=$cos$(\widehat{Q_1,Q_2})$,
\begin{eqnarray}\label{off-shell pi LbL 2}
 F_1(Q_1^2,Q_2^2,t)\,=\,\mathcal{F}_{\pi\gamma\gamma}(-Q_1^2,-Q_3^2,-Q_2^2)\,\mathcal{F}_{\pi\gamma\gamma}(-Q_2^2,0,-Q_2^2)\nonumber\\
 F_2(Q_1^2,Q_2^2,t)\,=\,\mathcal{F}_{\pi\gamma\gamma}(-Q_1^2,-Q_2^2,-Q_3^2)\,\mathcal{F}_{\pi\gamma\gamma}(-Q_3^2,0,-Q_3^2)\,,
\end{eqnarray}
and the integration kernels $I_1(Q_1,Q_2,t)$ and $I_2(Q_1,Q_2,t)$ can be found in Ref.~\cite{Jegerlehner:2009ry}.

\end{document}